\documentclass{aa}
\usepackage{natbib,psfig,graphicx,ulem}
\usepackage{txfonts}
\usepackage{soul,color}

\def\mathnew{\mathsurround=0pt}
\def\simov#1#2{\lower .5pt\vbox{\baselineskip0pt \lineskip-.5pt
\ialign{$\mathnew#1\hfil##\hfil$\crcr#2\crcr\sim\crcr}}}

\def\MeV{Me\kern-0.11em V}
\def\keV{ke\kern-0.11em V}

\begin{document}

\title{Orientations of very faint galaxies in the Coma cluster
\thanks{Based on observations obtained with MegaPrime/MegaCam, a joint project
of CFHT and CEA/DAPNIA, at the Canada-France-Hawaii Telescope (CFHT) which
is operated by the National Research Council (NRC) of Canada, the Institut
National des Sciences de l'Univers of the Centre National de la Recherche
Scientifique (CNRS) of France, and the University of Hawaii. This work is
 also partly based on data products produced at TERAPIX and the Canadian
Astronomy Data Centre as part of the Canada-France-Hawaii Telescope Legacy
Survey, a collaborative project of NRC and CNRS.
Based on observations obtained with MegaPrime/MegaCam, a joint project of CFHT
and CEA/DAPNIA, at the Canada-France-Hawaii Telescope (CFHT) which is operated
by the National Research Council (NRC) of Canada, the Institute National des
Sciences de l'Univers of the Centre National de la Recherche Scientifique of
France, and the University of Hawaii.}}

\author{C.~Adami \inst{1} \and 
R. Gavazzi \inst{2} \and
J.C. Cuillandre\inst{3} \and
F.~Durret \inst{2} \and
O.~Ilbert \inst{1} \and
A.~Mazure \inst{1}  \and
R. Pell\'o \inst{4} \and
M.P. Ulmer \inst{1,5} 
}

\institute{
LAM, P\^ole de l'Etoile Site de Ch\^ateau-Gombert,
38, rue Fr\'ed\'eric Joliot-Curie,
13388 Marseille Cedex 13, France
\and
Institut d'Astrophysique de Paris, CNRS, UMR 7095, Universit\'e Pierre et Marie Curie,
98bis Bd Arago, 75014 Paris, France
\and
Canada-France-Hawaii Telescope Corporation, Kamuela, HI 96743
\and
Laboratoire d'Astrophysique de Toulouse-Tarbes, Universit\'e de Toulouse,
CNRS, 14 Av. Edouard Belin, 
31400 Toulouse, France
\and
Department of Physics and Astronomy, Northwestern University,
2131 Sheridan Road, Evanston IL 60208-2900, USA
}

\date{Accepted . Received ; Draft printed: \today}

\authorrunning{Adami et al.}

\titlerunning{Orientations of very faint galaxy in the Coma cluster}

\sethlcolor{red}

\abstract
% context heading (optional)
{Models of large scale structure formation predict the 
%possible 
existence of preferential orientations for galaxies in clusters.}
% aims heading (mandatory)
{We have searched for preferential orientations of very faint galaxies
in the Coma cluster (down I$_{Vega}$$\sim$-11.5).}
% methods heading (mandatory)
{By applying a deconvolution method to deep u$^*$ and I band images
of the Coma cluster, we were able to recover orientations down to very
faint magnitudes, close to the faintest dwarf galaxies in the Coma
cluster.}
% results heading (mandatory)
{No preferential orientations are found in more than 95$\%$ of the
cluster area, and the brighter the galaxies, the fewer preferential
orientations we detect. The minor axes of late type galaxies are
radially oriented along a northeast~- southwest direction and are
oriented in a north~- south direction in the western X-ray
substructures. For early type galaxies, in the western 
regions showing significant preferential orientations, galaxy major axes 
tend to be oriented perpendicularly to the
north~- south direction. In the eastern significant region and close to
NGC~4889, galaxy major axes also tend to point toward the two cluster
dominant galaxies. In the southern significant regions, galaxy planes
tend to be tangential with respect to the clustercentric direction,
except close to ($\alpha$=194.8, $\delta$=27.65) where the orientation is
close to $-15$ deg. Early and late type galaxies do not have the same
behaviour regarding orientation.}
% conclusions
{Considering various models, we give an interpretation 
which can account for the existence of preferential orientations for
galaxies (e.g. the tidal torque model or interactions with the intra
cluster medium). Part of the orientations of the minor axes of late type
galaxies and of the major axes of early type galaxies can be explained
by the tidal torque model applied both to cosmological filaments and
local merging directions. Another part (close to NGC~4889) can be
accounted for by collimated infalls. For early type galaxies, an
additional region ($\alpha =194.8, \delta = 27.65$) shows orientations
that probably result from local processes involving induced star
formation.}  \keywords{galaxies: clusters: individual (Coma)}

\maketitle

\section{Introduction}

Large-scale structures are one of the major predictions of cosmological
models, which reproduce well the observed galaxy filamentary structures
and clusters (e.g. de Lapparent et al. 1986) are well reproduced by
numerical simulations in cold dark matter models.  Within these
structures (filaments and clusters) galaxies evolve following several
possible behaviors, influenced by the surrounding matter.  One
interesting characteristic of these galaxies linked with the surrounding
dark matter is their orientation (e.g. Aubert et al. 2004, Pereira et
al. 2008).  The origin of their angular momentum and the influence of
the large scale matter distribution on this angular momentum is still
widely debated.

In clusters, for example, it is generally admitted that the major axis of
the central galaxy tends to be aligned with the cluster axis and with
the main infall direction onto the cluster (e.g. Binggeli 1982, West
1989, Trevese et al. 1992, Fuller et al. 1999,  
Torlina et al. 2007). This implies collimated
infalls (e.g. West 1994, Dubinski 1998), 
linking the cluster and the central galaxy growth.

Another prediction of simulations is that the rotation axis of
disk galaxies in large scale structures is perpendicular to the minor
axis of the surrounding structure (e.g. Navarro et
al. 2004). Considering filaments, this means that disk galaxies would be
perpendicular to the filament axis (see also Kitzbichler et
al. 2003). This is related to the acquisition of angular momentum
through tidal torques (e.g. Peebles 1969, Coutts 1996, Navarro et al. 
2004). This model predicts a 
merging of infalling galaxies following an axis perpendicular to the 
infalling direction, inducing therefore a resulting angular momentum toward the 
infalling direction. 

%Assuming that the angular momentum is perpendicular to the
%major plane of the resulting galaxy, this implies that the major plane of the
%galaxy is perpendicular to the infalling direction.

Turbulence models (e.g. Shandarin 1974, Ozernoy 1978,
Efstathiou \& Silk 1983) predict coherent orientations of galaxies in
clusters. The planes of these galaxies should be parallel to the major
planes of the closest large scale structures (e.g. the closest
sheet of galaxies).  Another model (the hedgehog model, Godlowski et
al. 1998) predicts that the galaxy planes are perpendicular to the
direction pointing toward the cluster center, leading to a
spherical-like infall.

In such structures, primordial alignments could be rapidly erased by
strong dynamical interactions (e.g. Plionis et al. 2003) and it is not
clear whether or not we should detect preferential galaxy alignments in
clusters of galaxies. Galaxy orientation searches in clusters
(e.g. Hawley \& Peebles 1975) have
indeed led to a wide variety of results: some clusters show preferential
galaxy orientations (e.g. Kitzbichler et al. 2003, Plionis et al. 2003,
Aryal et al. 2007 and other Aryal et al. references therein), while some
others do not (e.g. Bukhari et al. 2003, Torlina et al. 2007).

Even when such alignments are detected, the results are not always
similar.  Studies can show alignments of galaxies
towards the dominant galaxy (e.g. Yang et al. 2006, Agustsson et
al. 2006), rotation axes along the direction of the surrounding large
scale structure (e.g.  Aryal et al.  2005a, Hu et al. 2006, Trujillo et
al. 2006), or rotation axes perpendicular to the direction of the
surrounding large scale structure (e.g. Aryal et al. 2005b).

The presence or not of preferential alignments and their direction seems
to be related to the cluster type (see Baier et al. 2003 or Aryal et
al. 2007). It also probably depends on the galaxy types (Aryal et al.
2005b) and magnitudes (Torlina et al. 2007), showing the need for a well
defined sample and for a galaxy type determination before searching for
preferential orientations. It is also highly desirable to have the
information on the large scale structures surrounding the cluster and on
the cluster building history. All these data are available only for a
few clusters and we chose to concentrate here on Coma.

The Coma cluster has been widely studied in the literature (see
e.g. Biviano et al. 1998 for a review), including galaxy orientation
determinations. For example, Wu et al. (1997) found parallel
orientations of part of the Coma spiral galaxy disks with the large
scale filament joining Abell~1367 and the Coma cluster. Kitzbichler et
al. (2003) detected preferential alignments for Coma galaxies located in
substructures, while Torlina et al. (2007) found no significant
alignments except for the dominant galaxies of the cluster.  However,
these studies were limited to relatively bright galaxies and we want to
investigate in the present paper the faint galaxy behavior, because low
mass galaxies could have been formed in halos around larger galaxies and
in this case should not show any preferred alignment on a scale beyond
that of the giant galaxies.  For this, we analyzed a large set of
galaxies in the Coma cluster, correlating the orientation information
with the multicolor type of the galaxies (estimated using broad band
spectral fits and called in the following spectromorphological type), the location of 
the X-ray gas (Neumann et al. 2003) and
the known infall directions around this cluster (Adami et al. 2005).  We
also dedicated special care to the orientation estimate, using a
deconvolution technique and a comparison with HST-ACS data.

The redshift of Coma is 0.023, and we will assume a distance to
Coma of 100 Mpc, H$_0$ = 70 km s$^{-1}$ Mpc$^{-1}$, $\Omega _\Lambda =0.7$,
$\Omega _m =0.3$, distance modulus = 35.00. The scale is 0.46
kpc arcsec$^{-1}$. All magnitudes are given in the Vega system.

\begin{figure*} \centering
\mbox{\psfig{figure=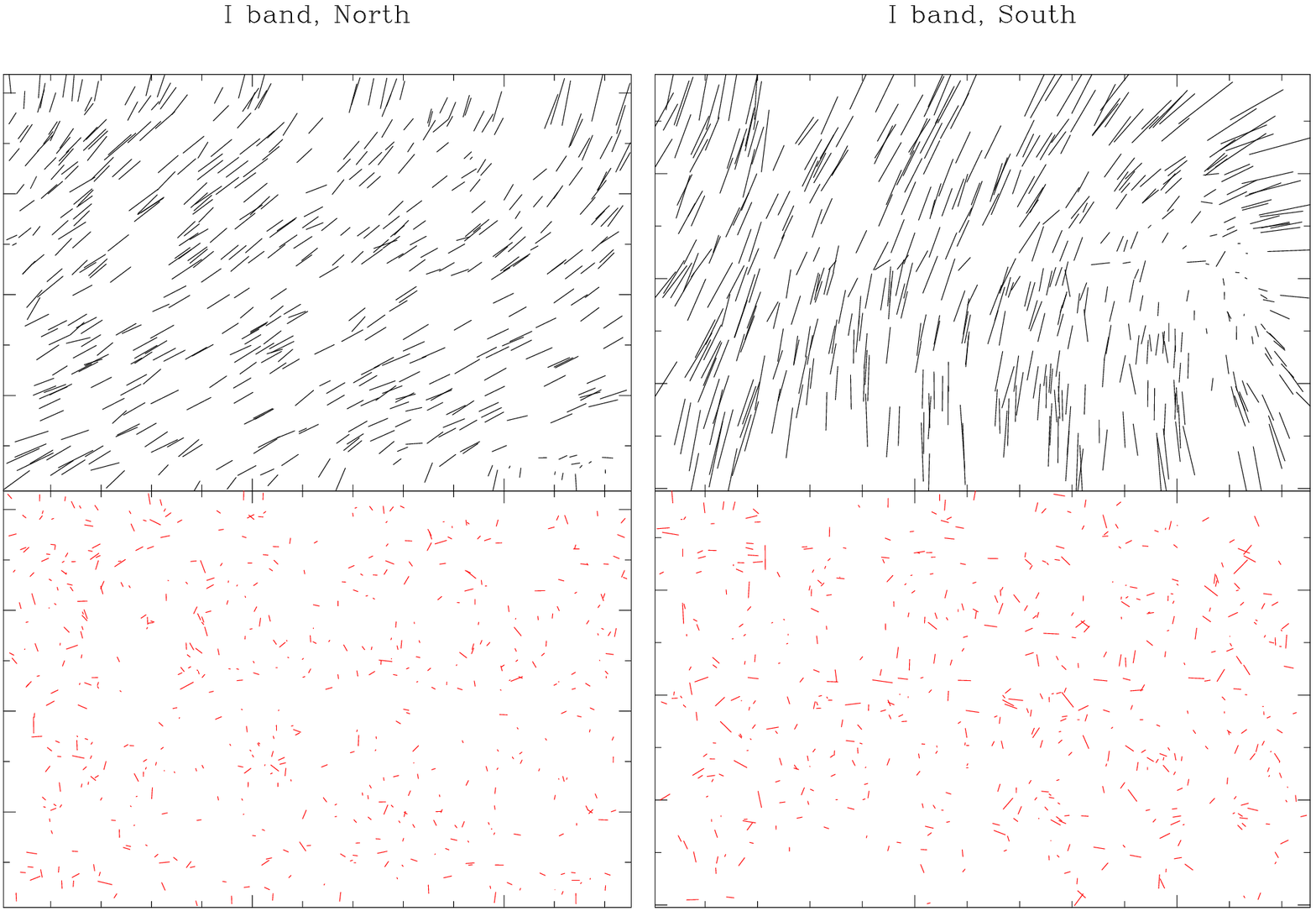,width=16cm,angle=0}}
\mbox{\psfig{figure=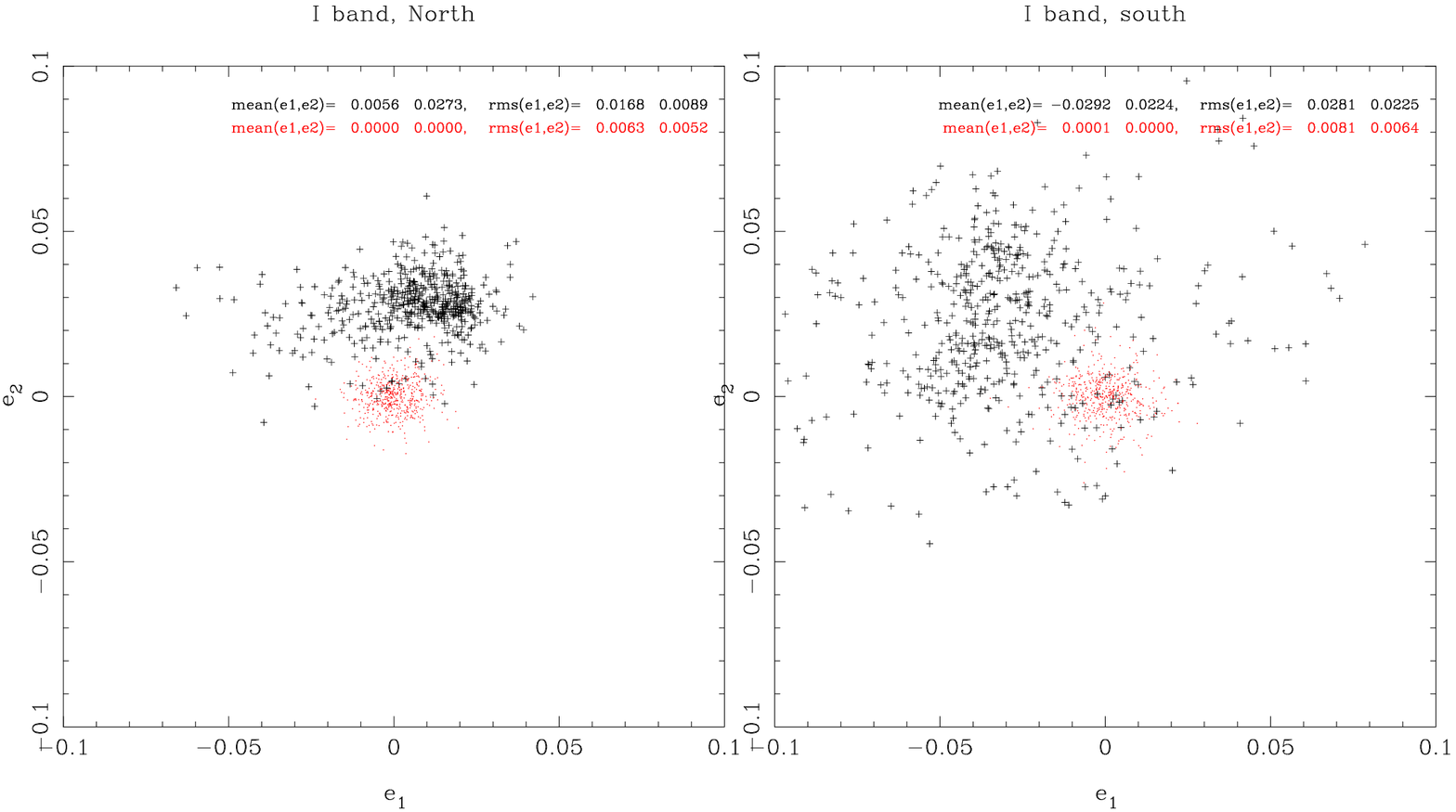,width=16cm,angle=0}} \caption[]{Shape of
stars before (top graphs) and after (middle graphs) PSF correction for
CFH12K I data. As usually done in weak lensing studies, ellipticities
given in bottom graphs are defined as complex numbers $e =
\frac{a-b}{a+b} \exp(2i \phi) \equiv e_1 + i e_2$.  Hence a positive
(resp. negative) $e_1$ corresponds to an object horizontally
(resp. vertically) aligned whereas $e_2$ codes for elongations along the
$\pm45^\circ$ directions. A circular object has $e_1=e_2=0$ as achieved
after PSF correction. Left graphs: north I band CFH12K data, right
graphs: south I band CFH12K data.}  \label{fig:corr_psf1}
\end{figure*}

\begin{figure}
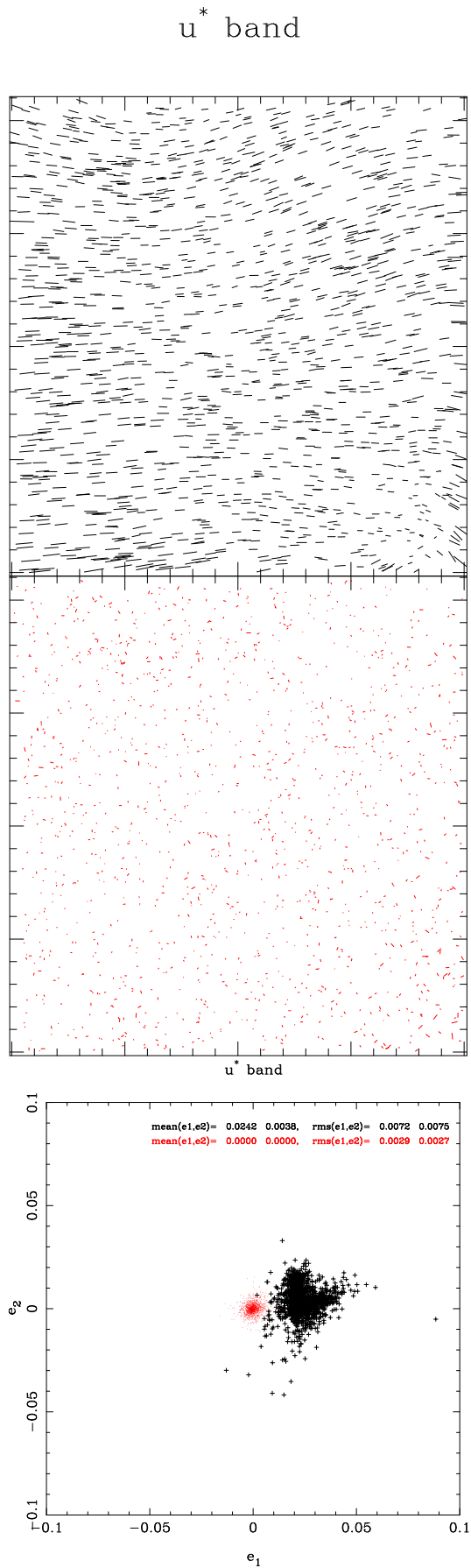
 \centering
\mbox{\psfig{figure=anis_u_1.ps,width=7cm,angle=0}}
\mbox{\psfig{figure=anis_u_2.ps,width=7cm,angle=0}}
\caption[]{Same as Fig.~\ref{fig:corr_psf1} for u* Megacam data.} 
\label{fig:corr_psf2} 
\end{figure}

\section{Data and methods}

\subsection{Imaging data}

Our data were obtained at the CFHT with the CFH12K and Megacam cameras
and were fully described by Adami et al. (2006) and Adami et
al. (2008), so we give here only a brief description.

The CFH12K B, V, R, and I images cover a common area of 42$\times $52~arcmin$^2$
centered on the two major galaxies of the Coma cluster (NGC~4874 and
NGC~4889). The images are made of two CFH12K pointings. The whole field
was observed in four bands (B, V, and I with similar depth). The R band
is slightly deeper and is complete down to R$\sim $24. We excluded
galaxies brighter than R=18, as well as all the areas located within 2
times the radius of R$\leq$18 galaxies, in order to avoid saturation
effects. This excludes in particular the regions very close to the two
major galaxies.

The Megacam u$^*$ image covers a 1$\times $1 deg$^2$ field in a single
pointing, encompassing the CFH12K field of view. It is $\sim$90$\%$ complete
approximately at the same magnitude as the R band image.

In order to compute orientations for faint galaxies, we chose to limit
our analysis to the two extreme bands: u$^*$ (with a seeing of 1.1'' and 
which had the most homogeneous data reduction because of the single field 
observation) and I (which had the best seeing: 0.8'' for the south and 1'' for
the north data). This allowed us also to investigate the
wavelength dependence of the orientation determination.

\subsection{Preliminary data analysis and possible biases}

Before trying to compute galaxy orientations it is crucial to estimate
cluster membership. Nearly all published studies are based on
spectroscopic information and are therefore limited to relatively bright
magnitudes. Here, we want to study galaxies as faint as M$_R \sim
-10$. This is obviously much too faint to allow a spectroscopic redshift
measurement for the whole galaxy sample. We therefore chose to apply the
photometric redshift techniques described in Ilbert et al. (2006) and
Bolzonella et al. (2000), based on the spectral fitting of synthetic
templates on the u$^*$, B, V, R, and I data. This allowed us to
discriminate between galaxies at redshift greater and lower than 0.2
along the Coma cluster line of sight. We refer the reader to Adami et
al. (2008) for a more detailed discussion on the accuracy of redshift
estimates based on the comparison with spectroscopic redshifts.

Of course, this does not mean we completely avoid including in our
sample galaxies that are at z$\leq$0.2 (in the field or in loose groups)
but are not part of the Coma cluster.

If we consider the field galaxy luminosity function computed by Ilbert
et al. (2006) from similar data, we find that the field contribution
represents about 15$\%$ of the Coma cluster galaxies down to R=24. This
contribution is spread over the whole field and will act as a background
contribution of randomly oriented galaxies.  We could argue that
filaments at z$\leq$0.2 could provide preferential orientations for
their galaxy populations. However, even taking the largest known cosmic
bubble sizes, at least a dozen of such bubbles are superimposed between
Coma and z=0.2, probably making the projected orientations close to
random.

Similarly, we estimated that the contribution of galaxies in loose
groups is concentrated in precise locations and
represents $\sim$5$\%$ of the Coma cluster galaxies down to R=24 (Adami et
al. 2008). Preferential orientations coinciding with these regions
are therefore  suspect, but we will show in the following that the
places where such orientations are detected do not coincide with the
positions of loose background groups at z$\leq$0.2.

The last possible bias comes from non Coma member galaxies at z$\leq$0.2
lensed by the Coma cluster mass. The redshift interval where galaxies are the
most likely to be lensed is not z$\leq$0.2, but we still estimate that the lensing
amplitude in this redshift range is between 50 and 90$\%$ of the
predicted shear for z$\sim$1 galaxies. However, the amplitude of the
shear remains moderate. It adds a tangential ellipticity (defined here as the
minor to major axis ratio) ranging from 0.05 for objects very close to the
cluster centre ($\sim$1' from the center) to less than 0.01 in the field
edges. This means that for background galaxies, lensing will slightly increase our ability to
detect tangential orientations while it will slightly lower our ability
to detect radial orientations.

In the cluster center, this will have a strong effect only on galaxies
that are not part of Coma (15$\%$ of field galaxies + 5$\%$ of loose
group galaxies) with an axis ratio higher than 0.95 (less than 20$\%$
of the total sample). So lensing will affect only (15$\%$ +
5$\%$)$\times$20$\%$ $\sim$5$\%$ of the total sample in the cluster
center.

In the cluster outskirts, lensing will affect only the galaxies with an
axis ratio higher than 0.99 (less than 2$\%$ of the total sample), so
less than (15$\%$ + 5$\%$)$\times$2$\%$ $\sim$1$\%$ of the total sample
in the cluster outskirts.

\subsection{PSF deconvolution}

\begin{figure*} \centering
\mbox{\psfig{figure=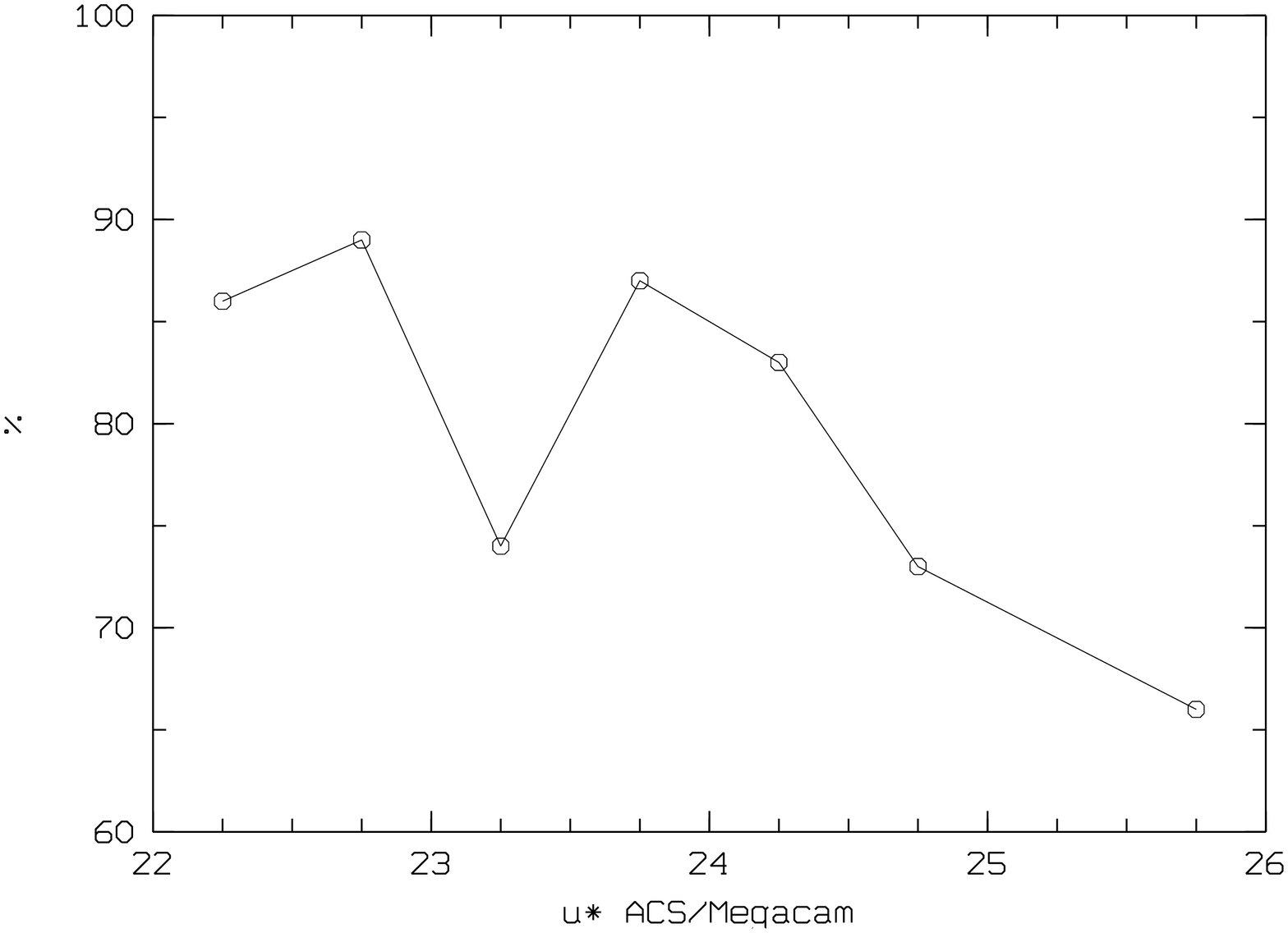,height=4cm,angle=270}\psfig{figure=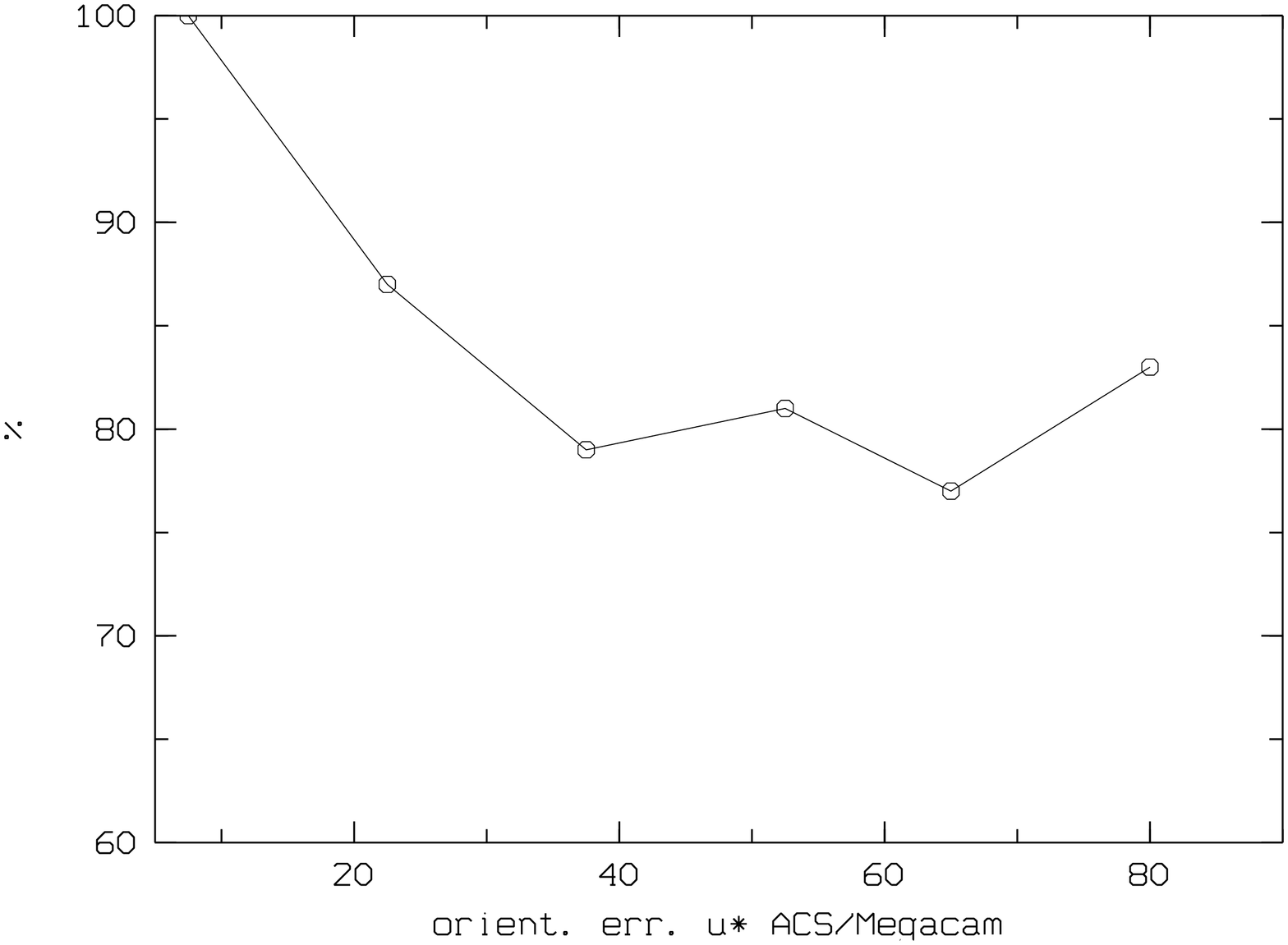,height=4cm,angle=270}\psfig{figure=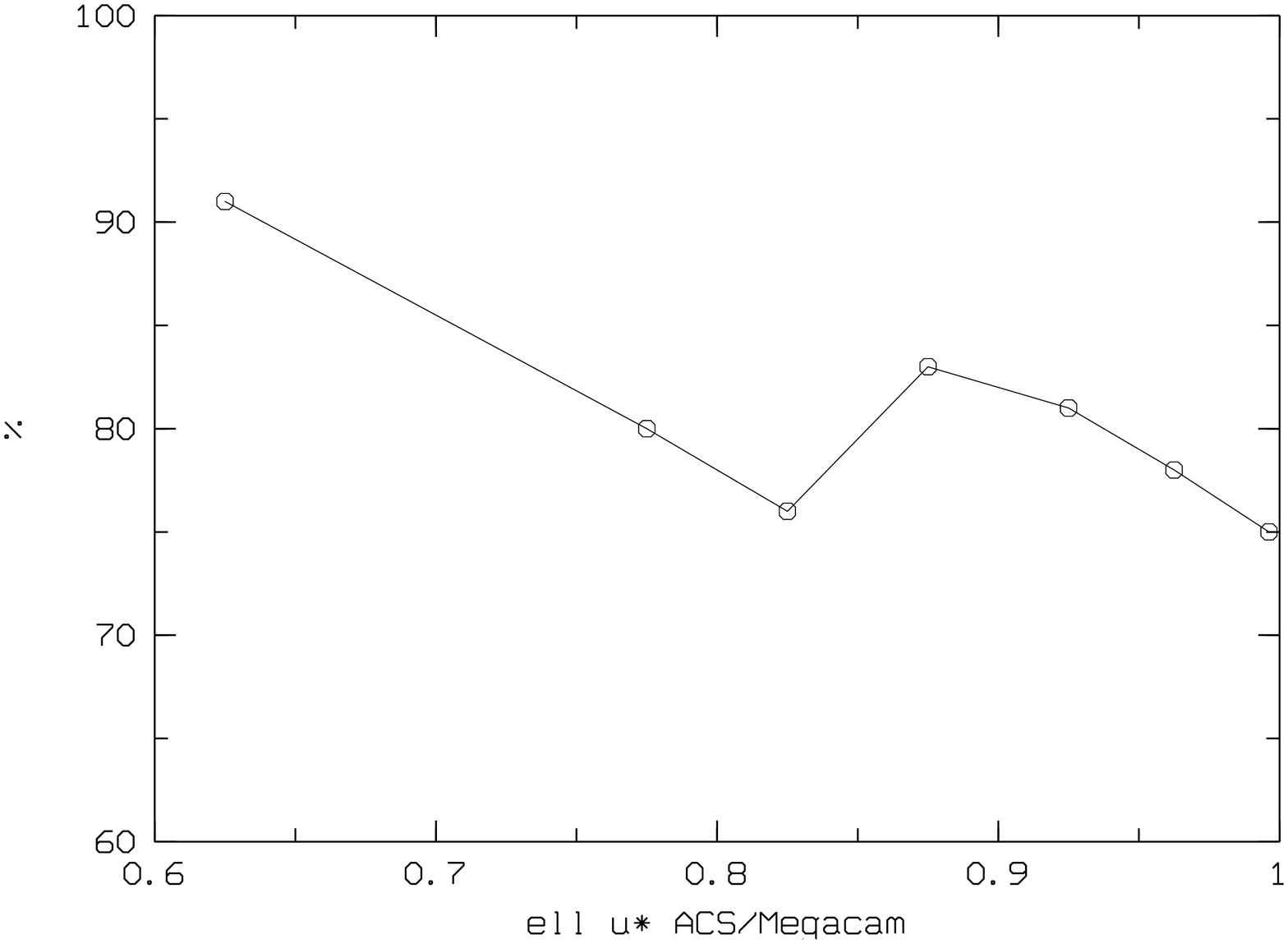,height=4cm,angle=270}}
\mbox{\psfig{figure=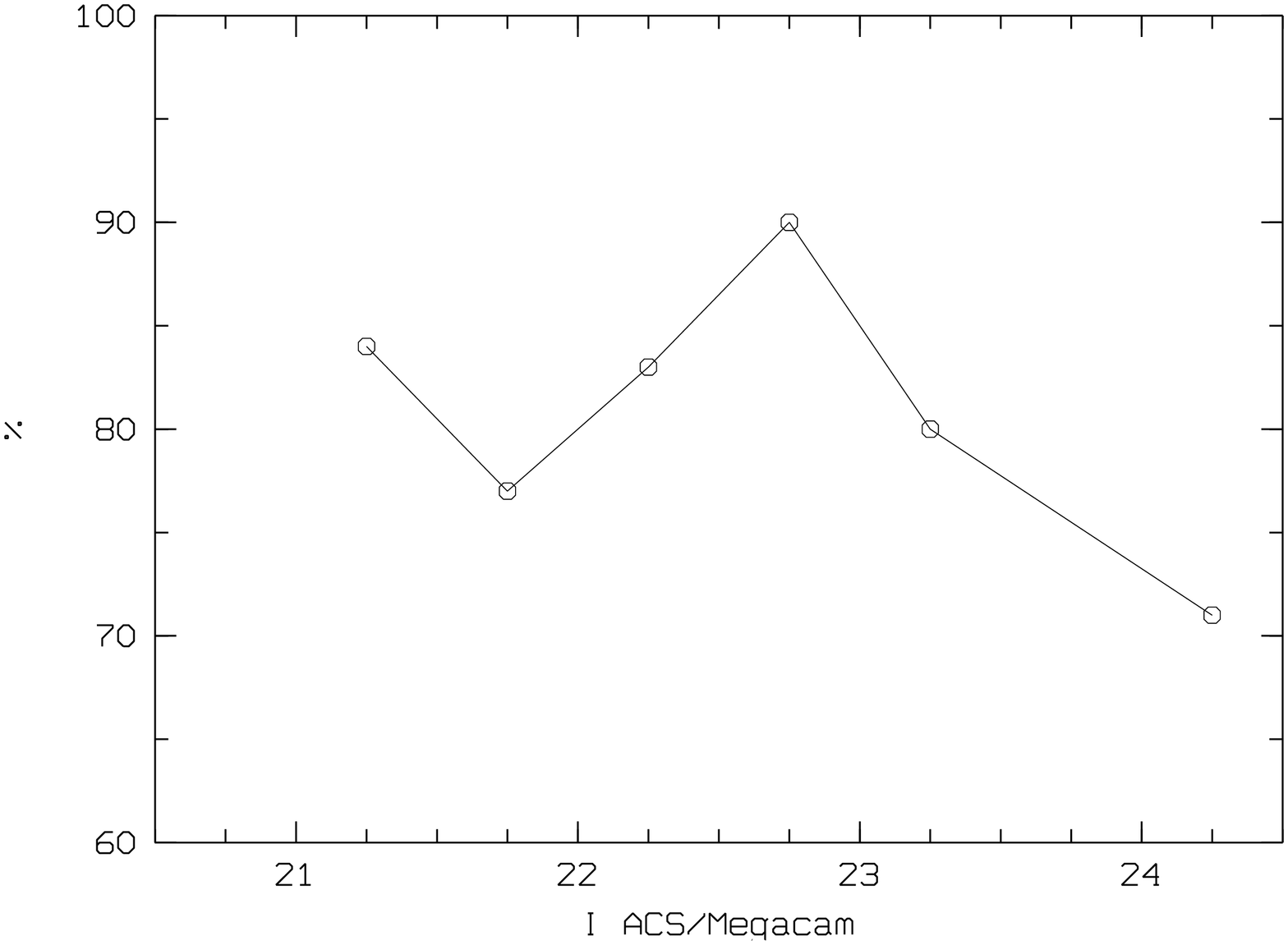,height=4cm,angle=270}\psfig{figure=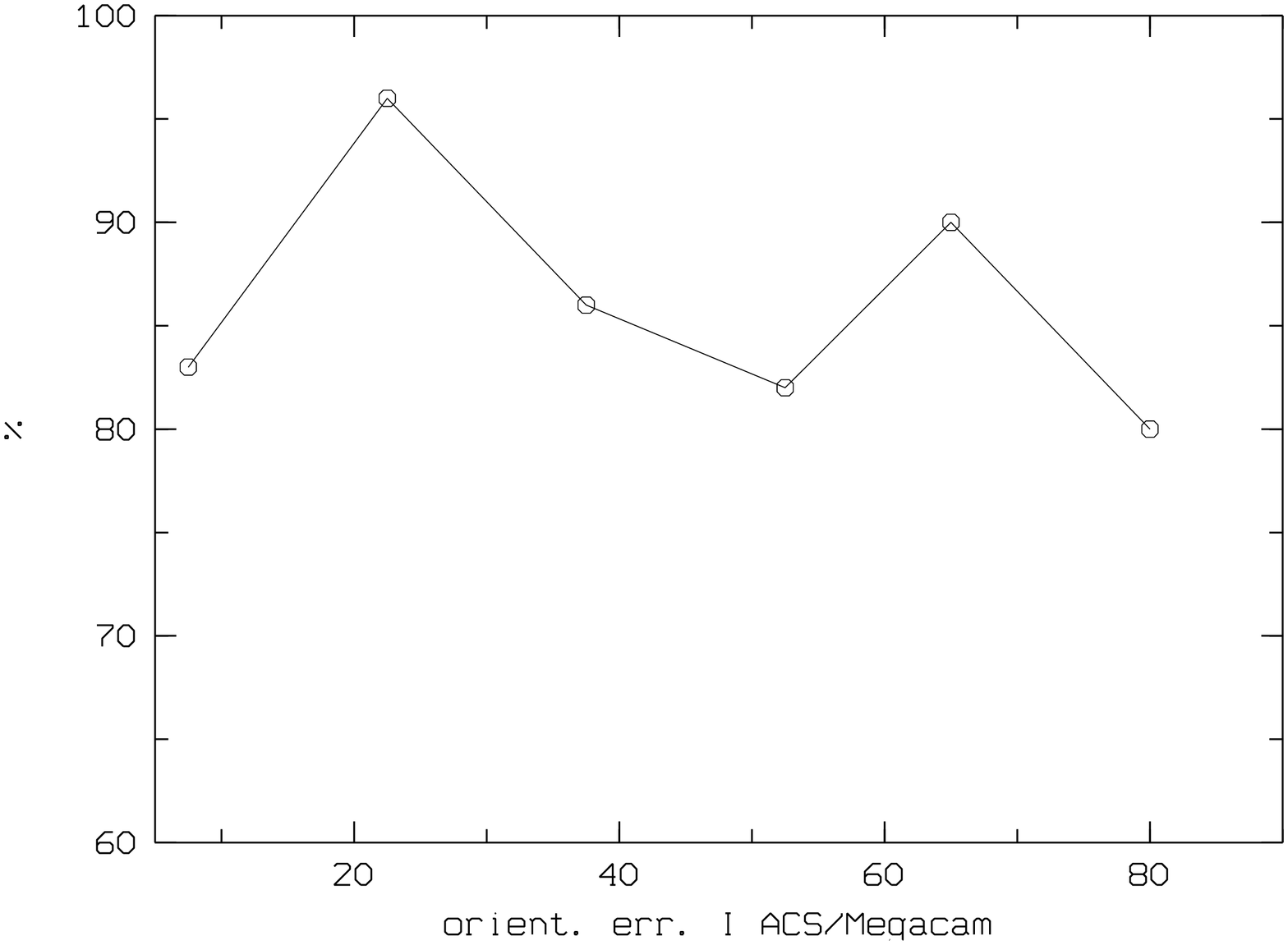,height=4cm,angle=270}\psfig{figure=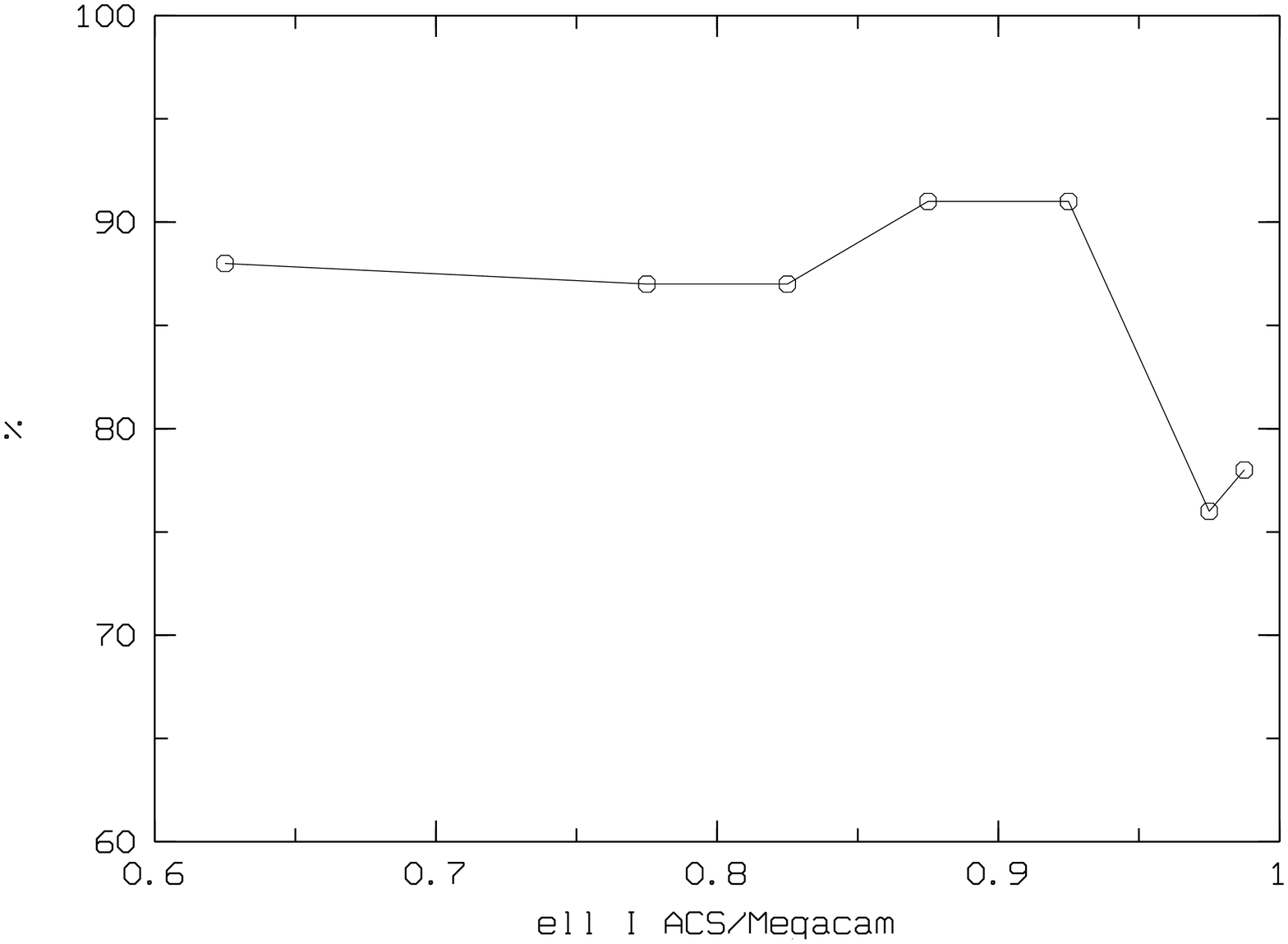,height=4cm,angle=270}}
\caption[]{Comparison of the orientations estimated in our data with our
deconvolution method and in HST-ACS data. Upper figures: Megacam u$^*$
and HST-ACS F475W data. Lower figures: I CFH12K data and HST-ACS
F814W. The numbers give the percentages of galaxies detected in our data
and in HST-ACS data with a difference in orientation estimate of less
than $\pm$26~deg as a function of (from left to right) magnitude in our
data, orientation uncertainty (in degrees) and minor/major axis ratio. }
\label{fig:compar}
\end{figure*}

\begin{figure*} \centering
\mbox{\psfig{figure=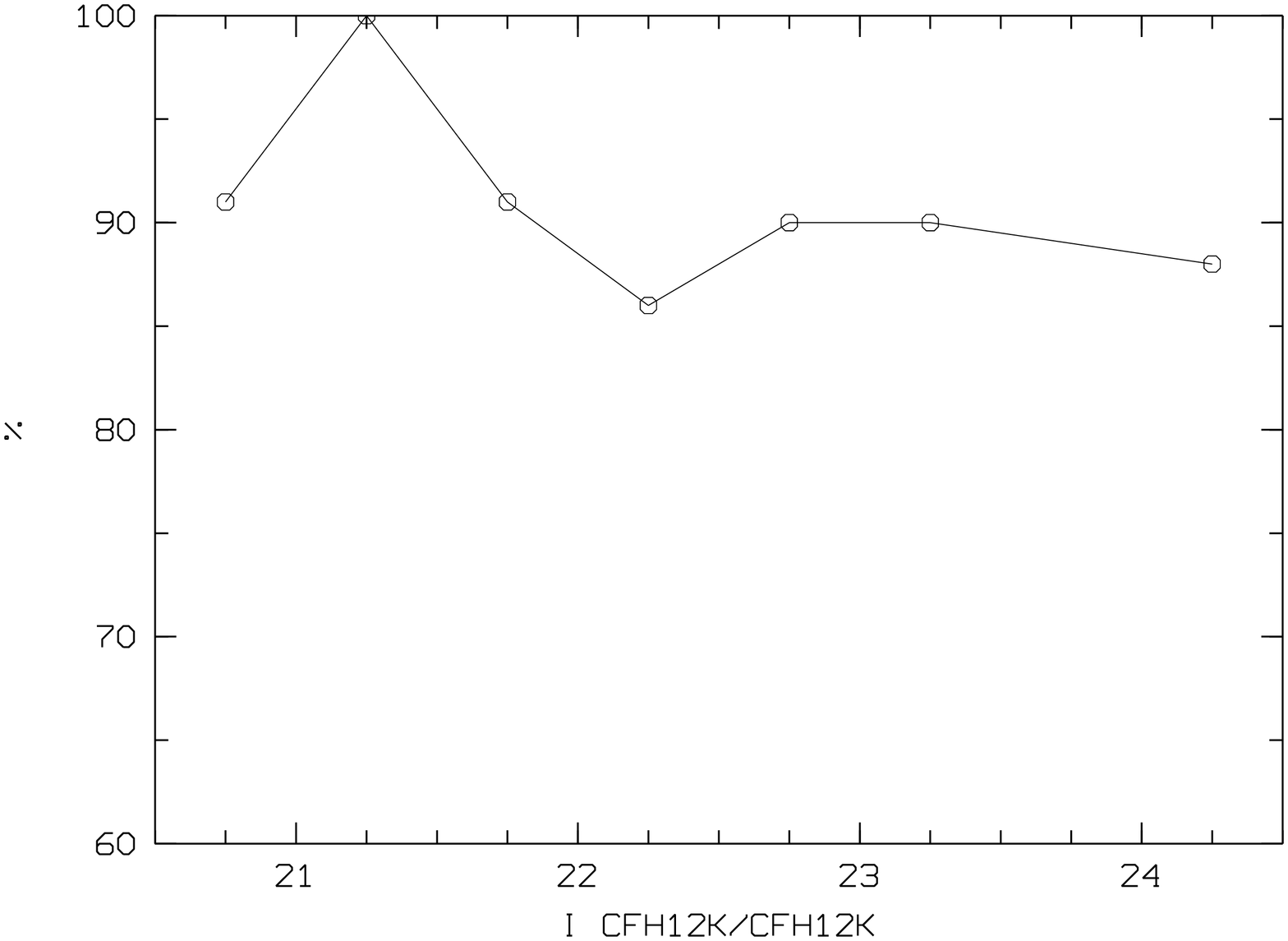,height=4cm,angle=270}\psfig{figure=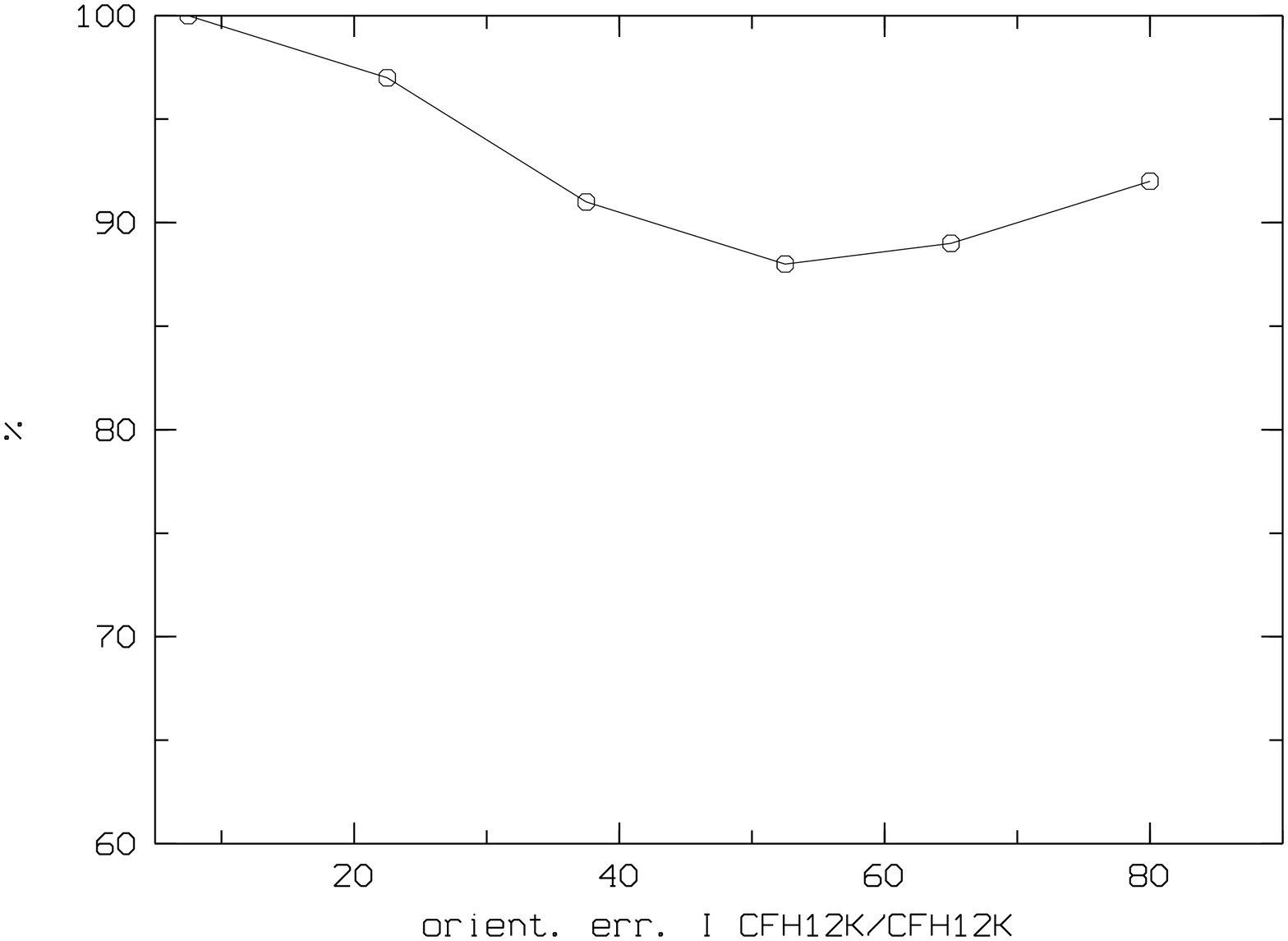,height=4cm,angle=270}\psfig{figure=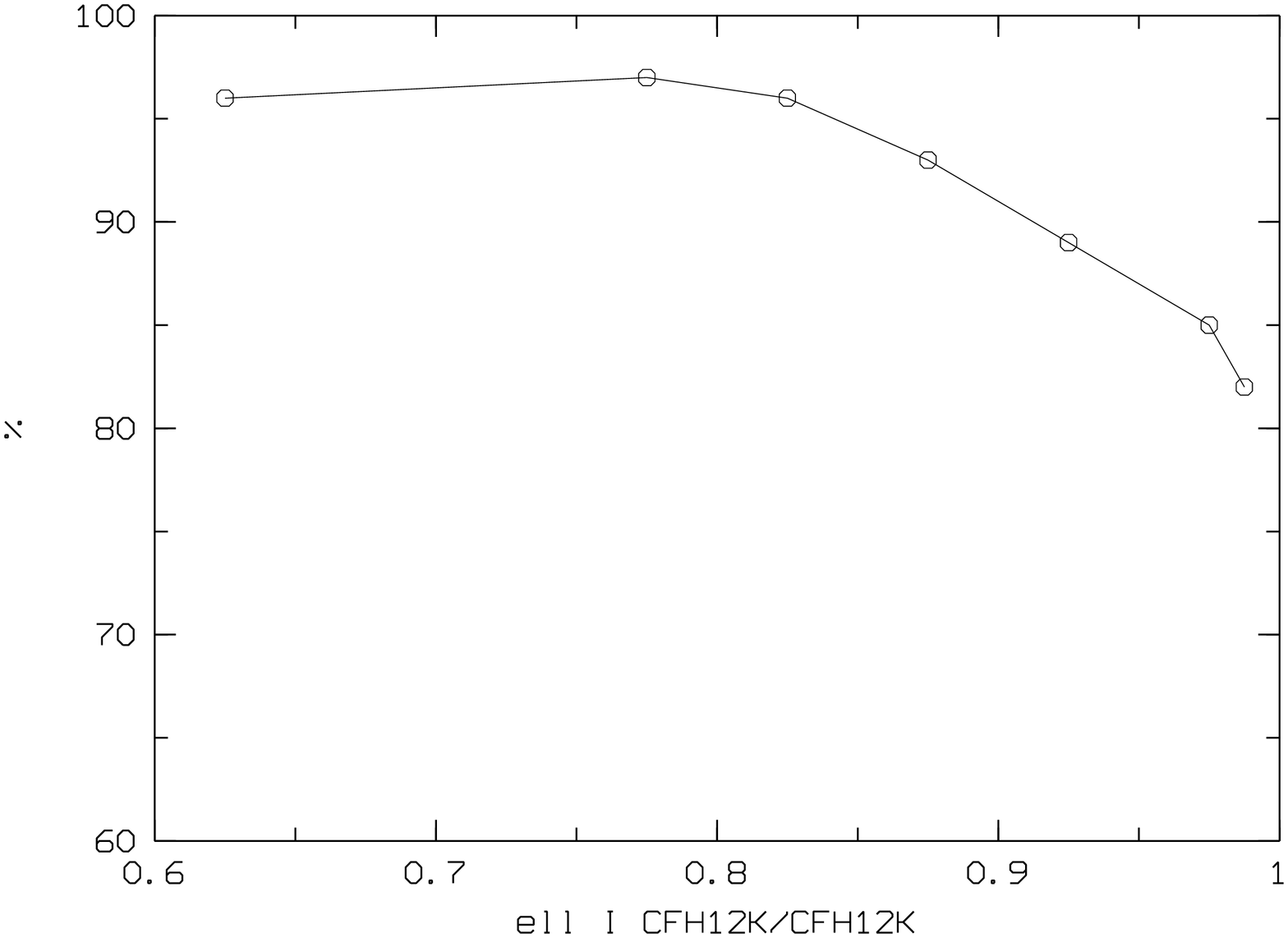,height=4cm,angle=270}}
\caption[]{Comparison of the orientations estimated in our CFH12K data
for galaxies observed both in the northern and southern fields. The
figures give the percentages of galaxies with a difference of
orientation estimate smaller than $\pm$26~deg as a function of (from
left to right) I magnitude, orientation uncertainty (in degrees), and
minor/major axis ratio. }  \label{fig:compar2}
\end{figure*}

\begin{figure} \centering
\mbox{\psfig{figure=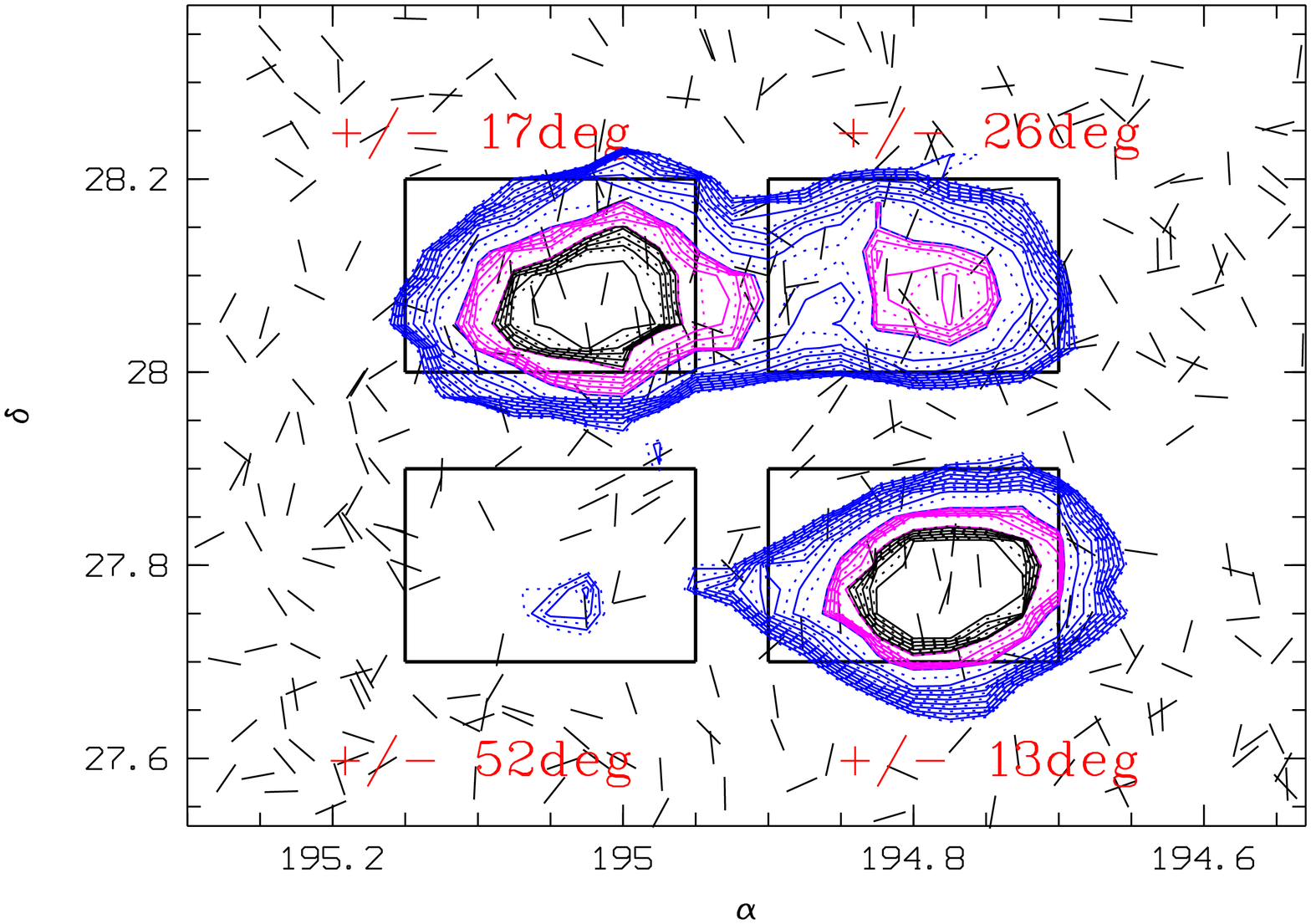,height=6cm,angle=270}}
\mbox{\psfig{figure=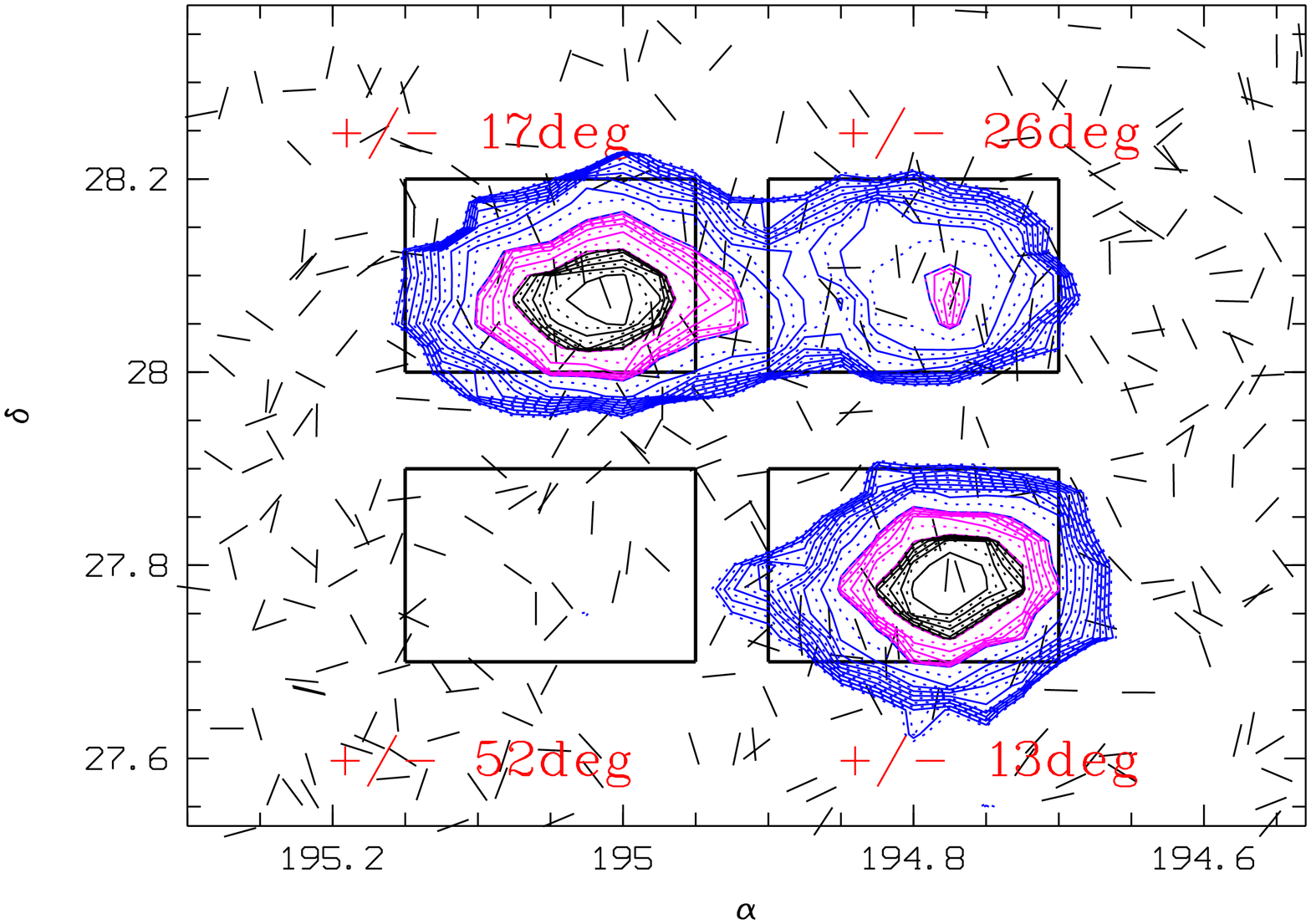,height=6cm,angle=270}}
\mbox{\psfig{figure=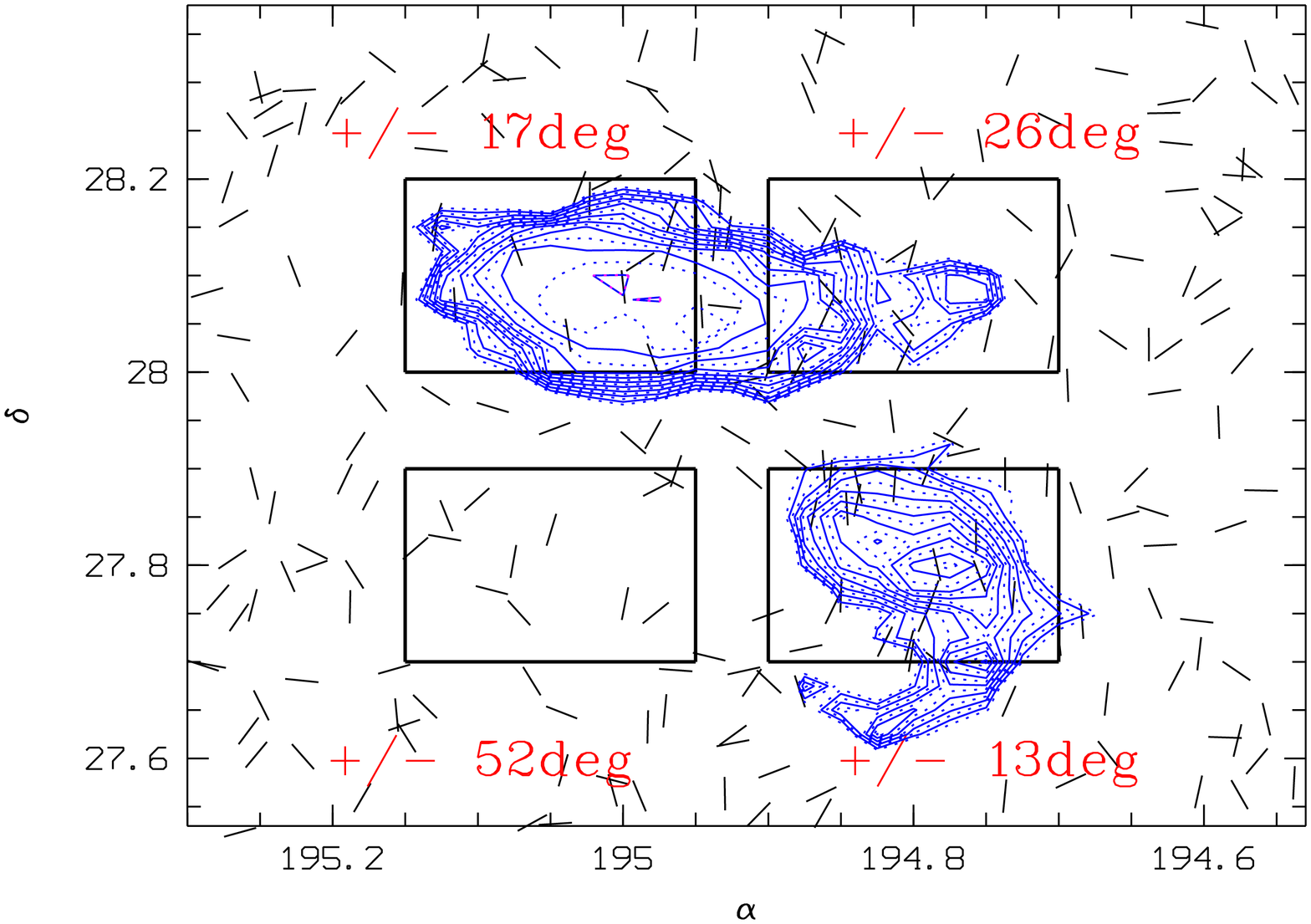,height=6cm,angle=270}}
\caption[]{Maps showing the simulated galaxy orientation distributions
and the four rectangular regions where we imposed a mean orientation of
0~deg and a given dispersion around this value (given in red
letters). Contours show the regions where the Kolmogorov-Smirnov test
detected significant alignments at the 88$\%$ (blue (heavy grey in black
and white version)), 99$\%$ (magenta (grey in black and white version)),
and 99.9$\%$ (black) levels, for the whole sample (top: 1 simulated
galaxy out of 10 is shown for clarity), for the spectromorphological
type 1 sample (middle: 1 simulated galaxy out of 8 is shown), and for
the spectromorphological types 2+3+4+5 (bottom: 1 simulated galaxy out
of 3 is shown). }  \label{fig:simul}
\end{figure}

Given the magnitude range we are considering, the galaxy sizes are
comparable to the seeing and therefore any attempt to measure shapes
will require a correction for smearing by the Point Spread Function
(PSF). The problem is very much the same as deconvolution for weak
gravitational lensing measurements for which several techniques have
been developed in the last 15 years.  Heymans et al. (2006) and Massey et
al. (2007), as part of the Shear Testing Program (STEP), proposed a
comprehensive quantitative study of the merits of the currently favored
methods. The one we implement here is built on the so-called KSB
techniques (Kaiser et al. 1995), further complemented by Hoekstra et
al. (1998). In this method, the shape (elongation, orientation) is
measured through weighted moments of the surface brightness distribution
of a given object on stars and galaxies. The deconvolution is performed
by requiring that stellar ellipticities have no preferential
orientation.  The blurring effect of the PSF is also
corrected for through the shear polarisability tensor $P^\gamma$ (see
Kaiser et al. 1995 for details). The PSF correction performed here was
previously applied to CFH12K data (Gavazzi et al. 2004) as well as
Megacam exposures of the CFHTLS deep survey (Gavazzi \& Soucail 2007)
with the aim of measuring weak gravitational lensing.  The strong
requirements usually imposed by weak lensing applications guarantee that
PSF smearing is properly corrected for in the sample of galaxies we are
considering. Although the main focus of the present work is the shape of
Coma member galaxies, a detailed study of the weak lensing shear signal
carried by background galaxies will be presented in a forthcoming paper
along with the details of the PSF correction (Gavazzi et al. 2008, in
preparation).  However, in order to give a brief view of the level of
control of systematic instrumental distorsions we show in
Figs.~\ref{fig:corr_psf1} and ~\ref{fig:corr_psf2} the shape of stars before and after PSF
correction.

%\subsection{Method}

%Given the magnitude range of our data, we deal with objects with sizes
%of the order of the seeing and of the order of the instrumental effects
%characterictic of the camera we used. It is therefore crucial to be able
%to deconvolve the images from instrumental effects before searching for
%preferential orientations.  The full method applied to our data is
%decribed in Gavazzi et al. (2008 in preparation) but we give here the
%salient points.

\subsection{Orientation validity}

We need to quantify the accuracy of the deconvolution method that we
apply to recover orientations. This method is supposed to give a
correct value for the orientation whatever the observation conditions
and the instrument.

We therefore first used HST-ACS data for a small subfield (public data
in the HST database, P.I. D. Carter) in order to compare non-deconvolved
orientation estimates with ours. The ACS orientations were
simply computed using the SExtractor package (Bertin $\&$ Arnouts
1996). Our estimates in the u$^*$ and I bands were compared to the F475W
and F814W HST filters respectively. We tested the influence of the
sample magnitude limit, of the projected ellipticity on the sky and of
the orientation uncertainty (given by our method) on the quality of our
orientation estimates (using CFH12K and Megacam data).

Results are given in Fig.~\ref{fig:compar} and we see that for the u$^*$
band, for u$^* \leq$25 and whatever the ellipticity and the orientation
uncertainty, we never have more than 30$\%$ of galaxies with
orientations differing by more than 26~deg from the ACS estimates (this
value of 26~deg has been chosen arbitrarily as the half of the width of a gaussian
function fit to a uniform orientation distribution between $-90$ and +90~deg). 
Considering
our whole sample and simply limiting the magnitude to u$^*$=25 leads to
87$\%$ of our u$^*$ orientation estimates within $\pm$26~deg of the ACS
estimates.

For the I band, 86$\%$ of our I orientation estimates are similar
to the ACS estimates for I$\leq$23.5.

However, the HST-ACS camera is not perfect, and the previous results are
an overestimate of the method uncertainties because including some
intrinsic dispersion due to HST-ACS camera instrumental effects. We
therefore also apply an internal comparison. If we consider the central
area of the I band image that was observed twice with the CFH12K camera,
we have two estimates of the orientations in this region, allowing to
compute the statistical difference. This method only gives a lower value
of the uncertainty due to the method, as it only tests the ground
observing conditions (that vary slightly between two observations) and
the CFH12K response inhomogeneities (the central field was not covered
by the same parts of the camera during the two observations). On the
other hand, this method does not test for example the general effect of
the atmosphere. Keeping this in mind, Fig.~\ref{fig:compar2} shows that
the difference between the two I orientation estimates is never greater
than 26~deg for more than 82$\%$ of the sample, whatever the magnitude,
the ellipticity, and the orientation uncertainty. For the whole sample,
we find that 91$\%$ of our I orientation estimates are similar in the
two I band observations.

We therefore conclude than about 90$\%$ (a good compromise between the
86, 87 and 91$\%$ previously estimated) of our orientations are correct
within $\pm$26~deg.

\section{Results}

In the following, orientations are counted counterclockwise from west. 
Each value is between  +90~deg and $-90$~deg.

\subsection{Global settings}

We search for preferential orientations of the galaxy axes with a
physical meaning: the major axis for early type galaxies (prolate like
objects) and the minor axis for late type galaxies (oblate like
objects).

We examine possible orientations of these axes 
in the u$^*$ and I bands. The u$^*$
band is more sensitive to recent star formation episodes and to recently
established preferential orientations while the I band is mainly
sensitive to the old star population and detects preferential
orientations established for a longer time.  We will concentrate on
galaxies fainter than I=18, beyond the usual limit for galaxy
orientation studies based on spectroscopic redshifts.  To our knowledge,
it is the first time that a multicolor approach is applied in such a
faint magnitude range.

Given the fact that preferential galaxy orientations seem to depend both
on galaxy type (e.g. Aryal et al. 2005b) and magnitude (Torlina et
al. 2007), we split our sample in two parts (type 1: elliptical
galaxies, and types 2+3+4+5: early spiral~+ late spiral+irregular+starburst
galaxies) based on the spectro-morphological types
from Ilbert et al. (2006). We also split the samples in three magnitude
bins for each of the two filters (I$\leq$23.5: faint I sample,
I$\leq$22: bright I sample, I$\leq$20: very bright I sample, u$^*
\leq$25: faint u* sample, u$^*\leq$23.5: bright u* sample, and
u$^*\leq$22: very bright u* sample).  We stress that our multicolor
types are based on a color classification in a 5 magnitude space and are
not real morphological types.

\subsection{Detection of significant preferential galaxy orientations}

The search for significant preferential galaxy orientations (whatever
the axis considered, major or minor) is made in several steps for a
given sample of galaxies (i.e. for a given spectro-morphological type
and a given magnitude bin).

- First, considering minor or major axes, 
we compute a smoothed map of the galaxy orientations of the
sample (the mean map in the following), using an adaptative kernel
technique (e.g. Adami et al. 1998a and references therein).  Each pixel
of the map is set to the mean orientation value at these
coordinates. The adaptative kernel technique represents a good balance
between the spatial resolution and the number of individual galaxies
considered to compute the mean value of the orientation in a given
pixel.

- Second, using a bootstrap technique with 1000 resamplings (for a given
pixel, this is equivalent to generating 1000 subsamples obtained by
removing one object from the total galaxy sample contributing to this
pixel) and a Kolmogorov-Smirnov test, we estimate in each pixel the mean
probability (over the 1000 resamplings) for the orientation distribution
in this pixel to be different from a completely random orientation
distribution (with the same number of galaxies). This allows to generate
maps of Kolmogorov-Smirnov probabilities to be different from random
orientation distributions.

- Third, we need to quantify the probability cutoff to apply to these
probability maps in order to recover the real signal (regions with
preferential orientations) without adding wrong detections (regions with
random galaxy orientations). The optimal probability cutoff proved to be
directly related to the number of galaxies available.  To assess this
point, we generated fake samples of galaxy orientations using the same
number of galaxies as in the real samples (within a given magnitude
range and spectromorphological type range). Orientations were randomly
distributed, except in 4 square regions where we imposed orientation
distributions with a mean value of 0 deg and dispersions of 13, 17, 26
and 52 degrees (52 degrees being the dispersion of a uniform orientation
distribution modelled by a gaussian function). In each of these 4 square
regions, we added 15$\%$ of randomly distributed field galaxies in order
to mimic the z$\leq$0.2 field galaxies which we have in our data and
which are not part of the Coma cluster. The goal is then to evaluate the
optimal probability cutoff to apply in order to recover the 13, 17, and
26 degree dispersion orientation distributions, without detecting the 52
degree one (which cannot be distinguished from a completely random
distribution using only the dispersion of the modelled gaussian
function). Fig.~\ref{fig:simul} shows a few examples of such
distributions with the recovered ones when using the optimal probability
cutoffs. These cutoffs are given in Table~\ref{tab:KMSP}. As a result,
we are able to optimally recover regions with preferential orientations
for the early spectromorphological types (types 1) and for the
spectromorphological types 2+3+4+5. 
%Spectromorphological types from 2 to
%5 taken individually are not numerous enough to allow detections.

\begin{table}
\caption{Optimal Kolmogorov-Smirnov probability cutoffs allowing to detect regions with real
  preferential orientations without adding wrong detections.}
\begin{center}
\begin{tabular}{lllll}
\hline
\hline
Sample & Bright & Faint \\ 
\hline
Whole & 83$\%$ & 88$\%$ \\ 
Types 1 & 82$\%$ & 83$\%$ \\ 
Types 2+3+4+5 & 82$\%$ & 82$\%$ \\ 
\hline
\end{tabular}
\end{center}
\label{tab:KMSP}
\end{table}

\subsection{Results for early spectro-morphological types.}

\begin{figure} \centering
\mbox{\psfig{figure=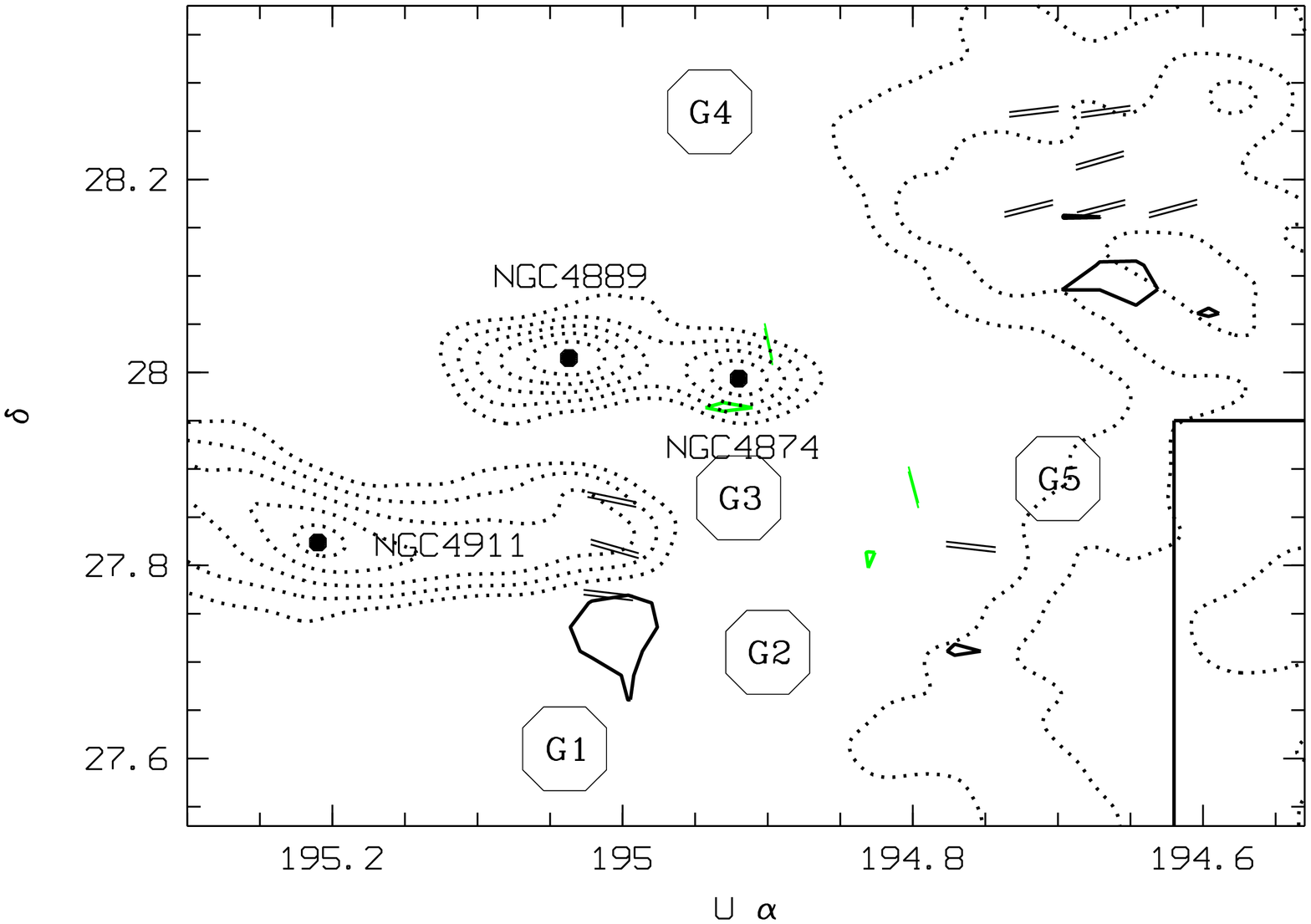,height=6.5cm,angle=270}}
\mbox{\psfig{figure=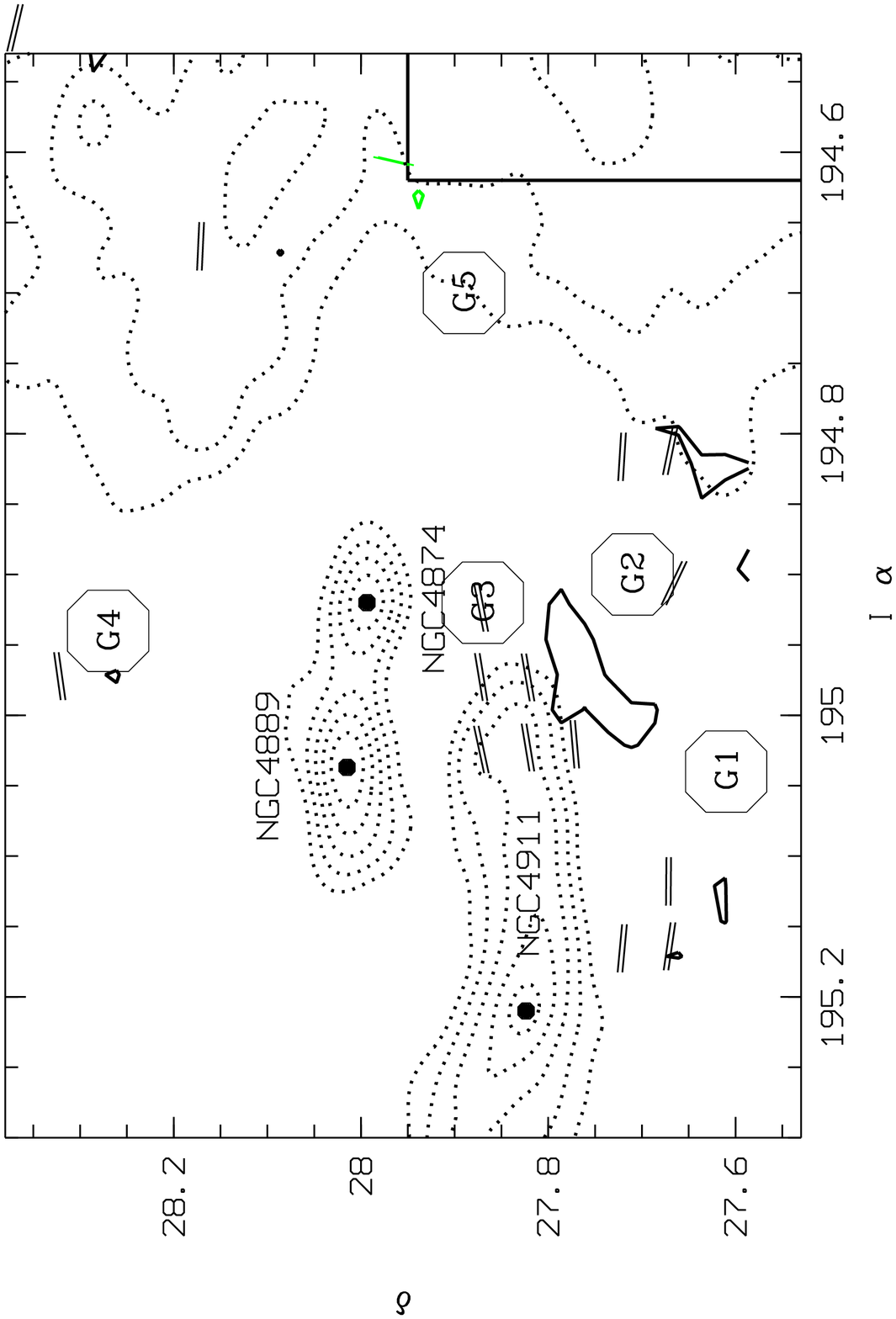,height=6.5cm,angle=270}} \caption[]{
Maps showing the regions where significant galaxy orientations are
detected for the whole galaxy sample. These figures show the minor axis
orientation for the late type galaxies and the major axis for the early
type galaxies. Upper figure: u* band, lower figure: I band. Dotted
contours: X-ray substructures from Neumann et al. (2003). Full contours:
areas where significant galaxy orientations are detected: black: types
1, green (grey in black and white): types 2+3+4+5. Octogons show the
loose z$\leq$0.2 background groups (see section 2.2).}
\label{fig:result1}
\end{figure}

\begin{figure} \centering
\mbox{\psfig{figure=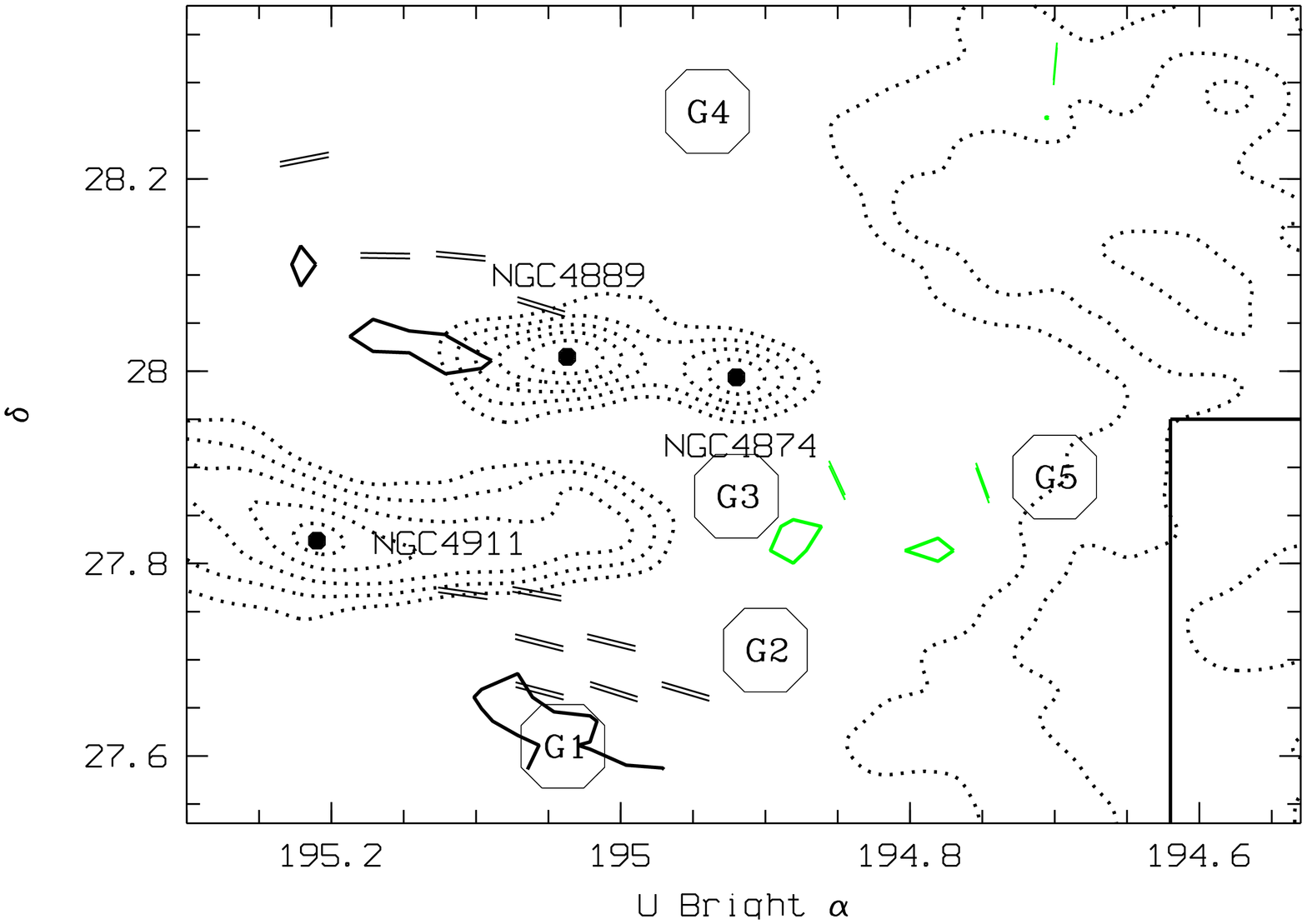,height=6.5cm,angle=270}}
\mbox{\psfig{figure=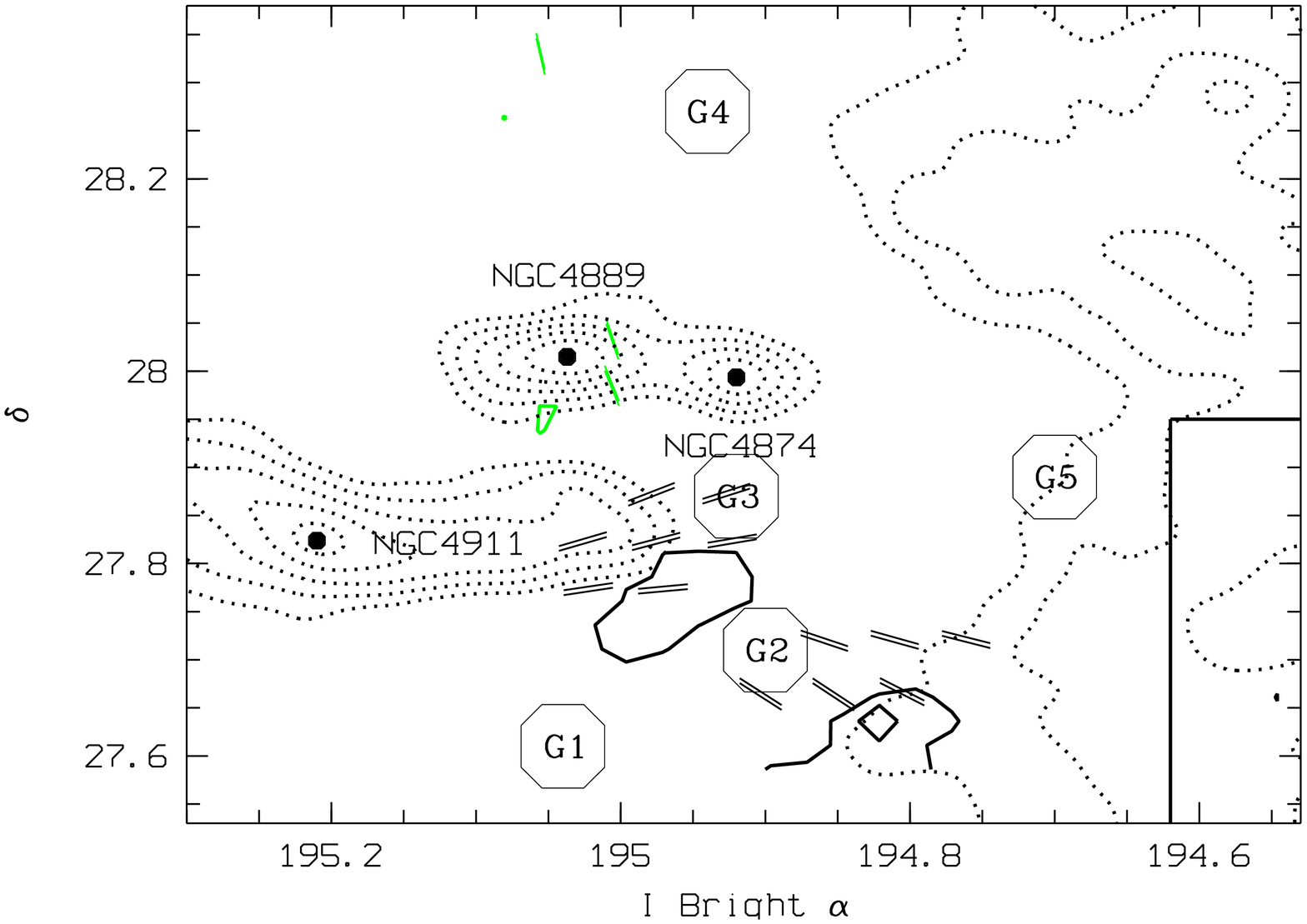,height=6.5cm,angle=270}}
\caption[]{Same as Fig.~\ref{fig:result1} for the bright galaxy samples.}  
\label{fig:result2}
\end{figure}

\begin{figure} \centering
\mbox{\psfig{figure=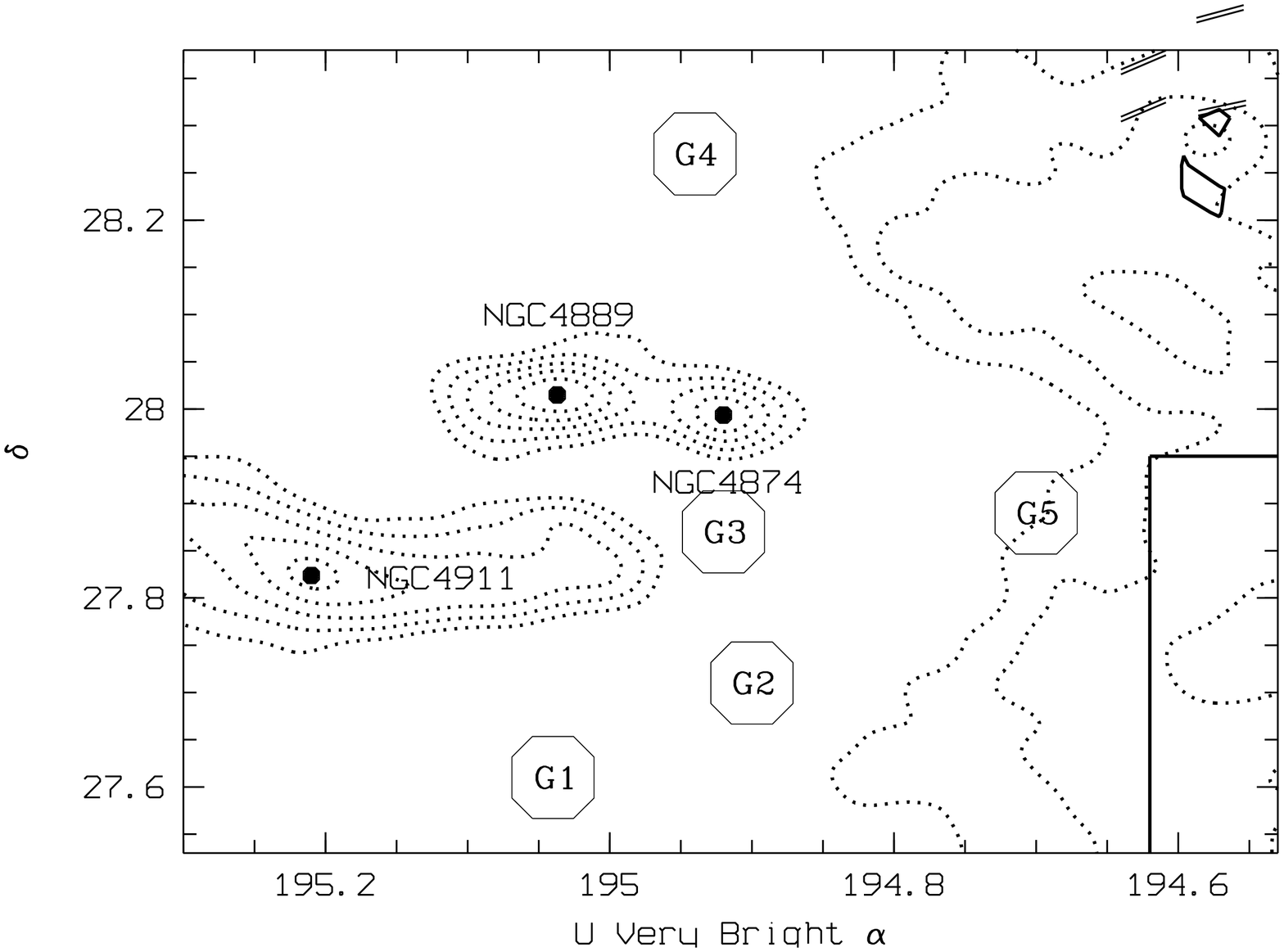,height=6.5cm,angle=270}}
\mbox{\psfig{figure=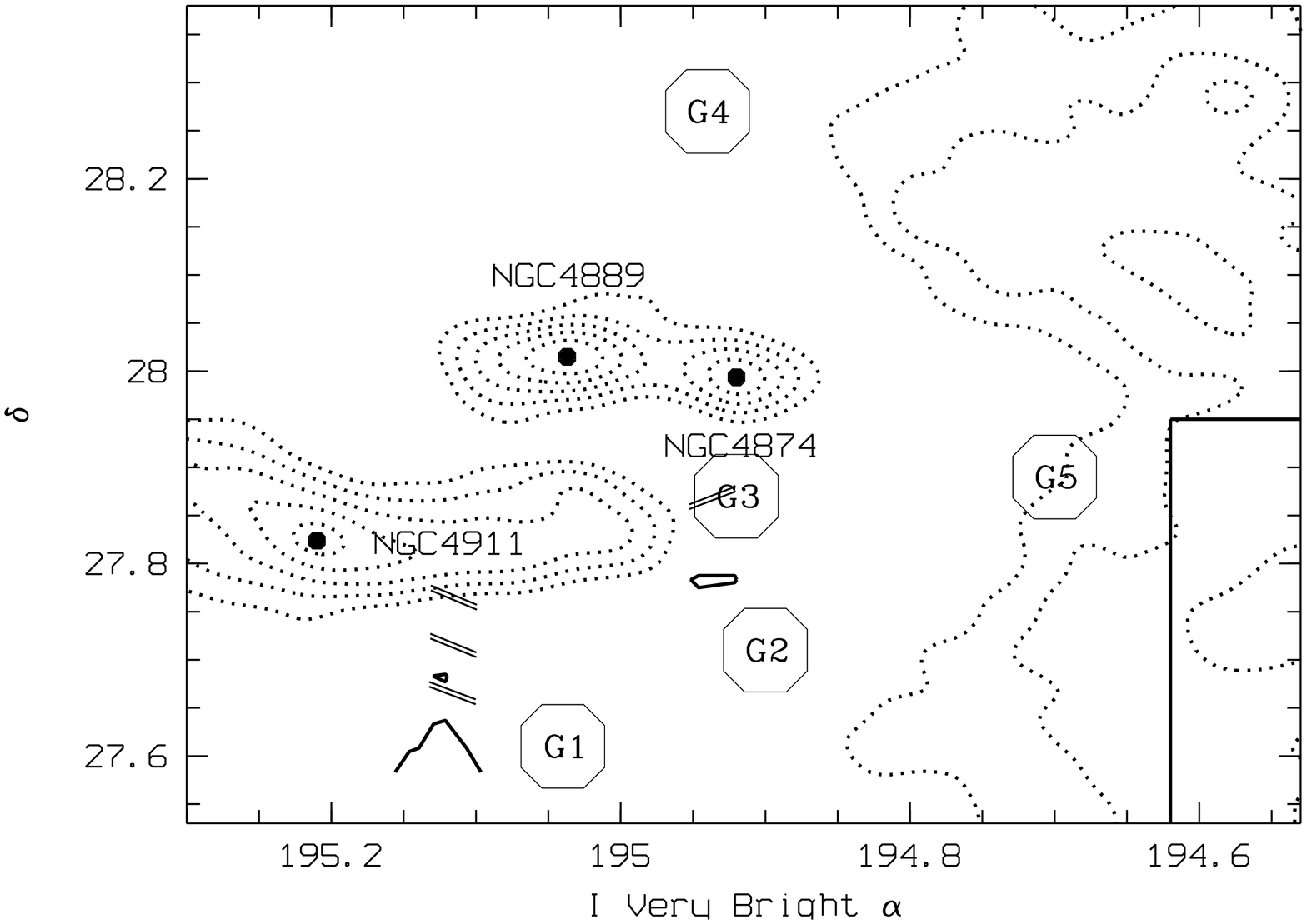,height=6.5cm,angle=270}}
\caption[]{Same as Fig.~\ref{fig:result1} for the very bright galaxy
samples. These samples are dominated by early type objects.}
\label{fig:result3}
\end{figure}

Results are displayed in Figures~\ref{fig:result1}, ~\ref{fig:result2},
and \ref{fig:result3}. These figures show the direction of the major
axis for early spectromorphological types. We only detect a few regions
with significant preferential orientations, mainly south of the cluster,
east of NGC~4889 and in the western X-ray substructures. In total, less
than 5$\%$ of the cluster area shows preferential orientations. The very
bright sample is the one that shows the least numerous regions with
significant preferential orientations. For this very bright sample, we
only tested the whole sample without discriminating into early and later
galaxy types because the number of available galaxies was too limited to
allow significant detections using the Kolmogorov-Smirnov method.
However, at I$<20$ the number of late type objects is limited ($<25$\%
from e.g. Adami et al. 1998b) and we can consider that the very
bright sample is largely dominated by early type objects.

\subsubsection{Western X-ray substructures and NGC~4889 vicinity.}

Some regions show radial orientations of the galaxy major axis relatively to
the clustercentric direction. This is mainly the case
in the western X-ray substructure regions (this trend is mainly detected
in the u* band data but is also marginally visible in the I band data
for the faintest galaxies), in the regions east of NGC~4889 (with an
orientation value close to the one provided by Kitzbichler et al. 2003
on a much brighter galaxy sample) and in the immediate vicinity of NGC~4889.

In the western regions (except for the very bright sample), the orientations
of the galaxy major axes are also perpendicular to the line 
joining the two X-ray infalling
groups detected in Neumann et al. (2003). This orientation does not
agree with the tidal torque model applied to the western (towards
Abell~779 and Abell~1367) or eastern (from Abell~2199) infalling
cosmological directions.  This orientation is probably related to the
presence of the two western infalling X-ray groups. A scenario
explaining this orientation would be the ongoing collapse of the two
infalling groups, imposing a north~- south infall direction to the tidal
torque model (e.g. Navarro et al. 2004).  In the eastern regions, only
the NGC~4911 infalling group can explain this radial orientation, if we
assume an ongoing merger of this group with the Coma cluster group
number 7 detected in Adami et al. (2005).

Close to NGC~4889, this radial orientation (similar to the direction
joining NGC~4889 and NGC~4874) can be explained by the collimated
infalls proposed by Torlina et al. (2007).

Collimated infalls applied to local mass concentrations can also predict
radial alignments of galaxy major axis toward these mass concentrations.
Using the X-ray substructures as tracers of these mass concentrations,
we could also interpret the orientations of the galaxies in the region
close to (194.95, 27.75) as a radial orientation relative to the
NGC~4911 location. However, we propose another explanation in
subsection 3.3.3., involving this time tangential orientations.

\subsubsection{The ($\alpha=194.95, \delta=27.75$) region.}

Several other regions in the south of the cluster show tangential galaxy
major axis orientations. For example, the orientations of the galaxies
in the region close to (194.95, 27.75) are tangential relative to the
clustercentric direction, and in good agreement with the tidal torque
model applied to a radial infall toward the Coma cluster center. This
trend appears for all I magnitude ranges and for the faint galaxy sample
in u*.

\subsubsection{The ($\alpha = 194.8, \delta=27.65$) region.}

The orientations of the major axes of the galaxies in the region close
to (194.8, 27.65) are neither tangential nor radial. This region located
at the south-east edge of the western X-ray subtructures shows
orientations of the galaxy major axes close to $-15$~deg that do not fit
with any grand design model (e.g. tidal torques or collimated infalls),
thus favouring a local process.

This region of preferential orientations only shows up in the I band
data.  Since the orientation of a given galaxy is traced by the
distribution of stars preferentially emitting in a given wavelength
interval, the u$^*$ star distribution will indicate the direction of the
recently formed stars while the I band will show the direction of the
old star distribution.  Assuming that star formation will be induced for
example by tidal forces, by dynamical friction, or by the intracluster
medium pressure, we could expect to have bursts of star formation toward
the direction of the motion (due to dynamical friction or to the
intracluster medium pressure effect). If these bursts are strong enough,
they can increase the luminosity of the galaxies along the motion
directions, and therefore modify the galaxy orientation estimates.  On
the one hand, our data do not show preferential orientations in this
region in the u$^*$ image; consequently, dynamical friction and the
pressure effect of the intracluster medium are not strong enough to
dominate the determination of the orientation. On the other hand, there
are preferential orientations detected in the I band; so intracluster
processes acting on galaxies must be strong enough to suppress in the
young star population the orientations seen in the old star population.

\subsubsection{Other regions.}

We also have tangential orientations in regions close to the G1 and G4
loose groups. However, this is perhaps not related to processes acting
inside the Coma cluster itself, as we do not know if the preferential
orientations are due to the Coma cluster or to the loose group galaxies.

\subsection{Results for late spectro-morphological types.}

We detect preferential orientations of the minor axis only in a few
regions. Along a northeast~- southwest direction minor axes show
radial orientations. The tidal torque model applied to the NGC~4839
cosmological infalling direction can explain this. In addition, we
detect north~- south orientations of minor axes in some places of the
western X-ray substructures. In this case, the tidal torque model needs
to be applied to a local merging direction joining the two components of
the X-ray substructure.

\section{Conclusions}

We have analysed deep optical images of the Coma cluster and searched
for preferential galaxy orientations, which are predicted by models
of cluster formation.  Our deconvolution method allows to recover valid
orientations down to very faint magnitudes, close to the faintest dwarf
galaxies in the Coma cluster.

Our main result is that more than 95$\%$ of the cluster area does not
show any preferential galaxy orientations, probably due to a rapid
isotropisation of the orientations. This negative result is fully
consistent with the recent findings of Torlina et al. (2007) based on a
H$\leq$14.5 galaxy sample.

Late spectromorphological type orientations can be explained by the
tidal torque model applied to the NGC~4839 infall direction and to a
north~- south merging direction between the two western substructures.

The occurence of regions showing significant preferential orientations for
early type objects seems to increase with decreasing luminosity of
the considered galaxy samples.  In the western regions with significant
galaxy orientations, galaxy major axes tend to be oriented
perpendicularly to the north~- south direction for early type galaxies and
galaxy minor axes tend to be oriented along the north~- south
direction for late type galaxies. In the eastern region with significant galaxy orientations
and close to NGC~4889 and NGC~4874, early type galaxy major axes also
tend to point toward the two cluster dominant galaxies. In the southern
regions with significant galaxy orientations, galaxy major axes of early
type galaxies tend to be tangential regarding the clustercentric
direction, except close to ($\alpha$=194.8, $\delta$=27.65) where the
orientation is close to $-15$~deg. This last region shows a different
behaviour in the u* and in the I band. Along a northeast~- southwest
direction, minor axes show radial orientations. The tidal torque model
applied to infall from cosmological filaments seems to apply only in the
southern regions. Applied to local merging processes, this model can
explain orientations detected in the western X-ray substructures and
east of NGC~4889. Finally, collimated infalls can account for
orientations detected in the immediate vicinity of NGC~4889.

Most of the regions where significant preferential orientations are
detected show different results for early or later type objects. This
illustrates the need for a spectral or morphological classification when
searching for preferential galaxy orientations. We have seen that a two
band strategy (the first one being sensitive to star forming galaxies
and the second to old stellar populations) also allows to infer
strategic informations such as ongoing group mergings or disentangling
old orientations imprinted in the old star population and recent star
formation bursts acting on the young star distribution.

\begin{acknowledgements}

{We thank the referee for constructive and useful comments. 
We acknowledge financial support from the GDR Galaxies, INSU-CNRS. 
The authors are grateful to the CFHT and Terapix (for the use of QFITS, SCAMP
and SWARP) teams, and to the French CNRS/PNG for financial support. MPU
also acknowledges support from NASA Illinois space grant
NGT5-40073 and from Northwestern University. }

\end{acknowledgements}

\end{document}